\documentclass[twocolumn,tighten,times,mathlines]{aastex631}

\usepackage{CJK}
\usepackage{amsmath}
\usepackage{booktabs}
\usepackage{longtable}  
\usepackage{float}
\usepackage{placeins} 

\submitjournal{ApJS}

\shorttitle{M Dwarf Parameters}
\shortauthors{Zhang et al.}

\graphicspath{{./}}

\begin{document}
\begin{CJK*}{UTF8}{gbsn}

\title{Half a Million M Dwarf Stars Characterized Using Domain-Adapted Spectral Analysis}

\correspondingauthor{Shuo Zhang \& Hua-Wei Zhang}
\email{szhang0808@pku.edu.cn; zhanghw@pku.edu.cn}

\author[0000-0003-1454-1636]{Shuo Zhang (张硕)}
\affiliation{Department of Astronomy, School of Physics, 
Peking University, Beijing 100871, P. R. China}
\affiliation{Kavli Institute of Astronomy and Astrophysics, 
Peking University, Beijing 100871, P. R. China}
\affiliation{Department of Astronomy, Tsinghua University, Beijing 100084, China}

\author[0000-0002-7727-1699]{Hua-Wei Zhang (张华伟)}
\affiliation{Department of Astronomy, School of Physics, 
Peking University, Beijing 100871, P. R. China}
\affiliation{Kavli Institute of Astronomy and Astrophysics, 
Peking University, Beijing 100871, P. R. China}

\author[0000-0001-5082-9536]{Yuan-Sen Ting (丁源森)}
\affiliation{Department of Astronomy, The Ohio State University, Columbus, OH, USA}
\affiliation{Center for Cosmology and AstroParticle Physics (CCAPP), The Ohio State University, Columbus, OH, USA}
\affiliation{Max-Planck-Institut f\"ur Astronomie, K\"onigstuhl 17, D-69117, Heidelberg, Germany}

\author[0000-0001-6767-2395]{Rui Wang (王瑞)}
\affiliation{Key Laboratory of Space Astronomy and Technology, National Astronomical Observatories, \\
Chinese Academy of Sciences, Beijing 100101, Peopleʼs Republic of China}

\author[0000-0003-3719-1990]{Teaghan O'Briain}
\affiliation{Department of Physics and Astronomy, 
University of Victoria, Victoria, BC V9E 2E7, Canada}

\author[0000-0003-0433-3665]{Hugh R. A. Jones}
\affiliation{Centre for Astrophysics Research, 
University of Hertfordshire, College Lane, Hatfield AL10 9AB, UK}

\author[0000-0002-8546-9128]{Derek Homeier}
\affiliation{Zentrum für Astronomie der Universität Heidelberg, 
Landessternwarte,  Königstuhl 12,  D-69117 Heidelberg,  Germany}

\author[0000-0001-7865-2648]{A-Li Luo (罗阿理)}
\affiliation{Key Laboratory of Optical Astronomy, 
National Astronomical Observatories, Chinese Academy of Sciences, 
Beijing 100012,  China}
\affiliation{University of Chinese Academy of Sciences, 
Beijing 100049, China}

\begin{abstract}
We present fundamental atmospheric parameters ($T_{\rm eff}$ and $\log g$) and metallicities ([M/H]) for 507,513 M dwarf stars using low-resolution spectra (R$\sim$1800) from LAMOST DR10. By employing Cycle-StarNet, an innovative domain adaptation approach, we successfully bridge the gap between theoretical PHOENIX synthetic spectra and observed LAMOST spectra, enabling parameter measurements even for lower signal-to-noise data (S/N$>$5). The fitting residual analysis shows a reduction from 2.0 times to 1.68 times the flux uncertainty. Comparing with available literature values, we find systematic offsets and precisions of 12$\pm$70 K in $T_{\rm eff}$, $-$0.04$\pm$0.17 dex in $\log g$, and $-$0.06$\pm$0.20 dex in [M/H]. The precision improves for higher quality spectra (S/N$>$50) to 47 K, 0.12 dex, and 0.14 dex respectively. The metallicity consistency between wide binaries shows a scatter of 0.24 dex, improving to 0.15 dex at S/N$>$50. We provide a comprehensive catalog including stellar parameters, spectral classifications, activity indicators, and binary/variability flags, establishing a resource for studies of the most numerous stellar population. The complete catalog can be accessed from url: https://doi.org/10.5281/zenodo.14030249.
\end{abstract}


\section{Introduction} \label{sec:intro}
M dwarf stars are the most common stars in our galaxy. They make up about 70\% of all stars \citep{1997AJ....113.2246R, 2010AJ....139.2679B} and 40\% of the stellar mass of the Milky Way \citep{2003PASP..115..763C}. These low-mass stars, found on the lower main sequence of the Hertzsprung-Russell (H-R) diagram, have masses below 0.7 M$_\odot$ \citep{2023Natur.613..460L}. They have exceptionally long lifespans and evolve very slowly \citep{2000nlod.book.....R}. This slow evolution helps them preserve the primordial information from their formation \citep{2022ApJ...927..122H}. Late-type M dwarf stars include brown dwarfs, which are important for understanding stellar formation \citep{2003AJ....126.2421C,2013ApJ...772...79A,2022A&A...660A..38W}.

M dwarf stars have become increasingly important in the search for exoplanets. Several potentially habitable planets have been discovered orbiting M dwarf stars, including Proxima Centauri \citep{2016A&A...596A.111R,2022A&A...658A.115F}, TRAPPIST-1 \citep{2017Natur.542..456G,2018A&A...613A..68G}, and LHS 1140 \citep{2020A&A...642A.121L}. This has created a growing need to better understand these stars' characteristics.

Spectroscopic analysis serves as a primary method for obtaining fundamental stellar atmospheric parameters, chemical abundance, radial velocity, activity, etc. Despite M dwarf stars' faintness, advances in telescopic equipment have enabled an increasing number of spectral samples to be collected and analyzed for these parameters. This includes not only high-resolution spectroscopic observations targeting the optical and infrared bands of M dwarf stars—often very nearby samples as highlighted by research from \cite{2013AJ....145...52M,2014A&A...564A..90R,2017ApJ...851...26V,2018A&A...620A.180R,2018A&A...610A..19R, 2020MNRAS.494.2718W}, and the Calar Alto high-Resolution search for M dwarf stars with Exo-earths with Near-infrared and optical Echelle Spectrographs (CARMENES; \citealt{2018A&A...612A..49R,2020SPIE11447E..3CQ}) series---but also encompasses medium to low-resolution large samples such as those by \cite{2013AJ....145..102L,2020AJ....159...30H, 2022ApJ...927..122H}.

Furthermore, numerous ongoing large-scale survey projects have provided valuable spectral data for M dwarf stars. High-resolution surveys include the near-infrared spectroscopic survey SDSS/APOGEE (R$\sim$22,500; \citealt{2017AJ....154...94M}) and GALactic Archaeology with HERMES (GALAH; \citealt{2021MNRAS.506..150B}). Medium to low-resolution surveys comprise the Large Sky Area Multi-Object Fiber Spectroscopic Telescope (LAMOST; \citealt{2012RAA....12.1197C}) with its medium-resolution survey (MRS; R$\sim$7500) and low-resolution survey (LRS; R$\sim$1800). Additional contributions come from the Milky Way Mapper of SDSS-V (\citealt{2017arXiv171103234K, 2023ApJS..267...44A}; R$\sim$2000) and Gaia Data Release 3 (DR3) XP spectroscopic observations (\citealt{2023A&A...674A...1G,2023A&A...674A...2D, 2023A&A...674A...3M,2023ApJS..267....8A}; R$\sim$15-85). These extensive datasets have significantly enhanced our understanding of these low-mass stars.

The urgent need to measure accurate parameters for the rapidly growing number of M dwarf spectra has led to substantial research efforts. For instance, using APOGEE spectra, \cite{2021A&A...649A.147S} measured spectroscopic parameters ($T_{\text{eff}}$, [M/H], $\log g$, $v_{mic}$) for 313 M dwarf stars. Additionally, \cite{2022ApJ...927..123S} derived individual chemical abundances for 14 elements in a sample of M dwarf stars, and \cite{2020ApJ...892...31B} presented a catalog of spectroscopic temperatures, metallicity, and spectral types for 5875 M dwarf stars. Regarding the vast LAMOST dataset of M dwarf spectra, several parameter measurement studies have been undertaken: \cite{2021ApJS..253...45L} implemented a data-driven approach to transfer labels (effective temperature and metallicity) from APOGEE to LAMOST low-resolution spectra, whereas \cite{2021RAA....21..202D} and \cite{2022ApJS..260...45D} measured fundamental parameters for M dwarf stars by fitting LAMOST DR7 low-resolution spectra to the PHOENIX model grid \citep{2012RSPTA.370.2765A,2013MSAIS..24..128A} and DR8 spectra to the MILES empirical spectral library \citep{2011A&A...532A..95F,2016A&A...585A..64S}.

\begin{figure*}[htb]
\centering
\includegraphics[width=135mm]{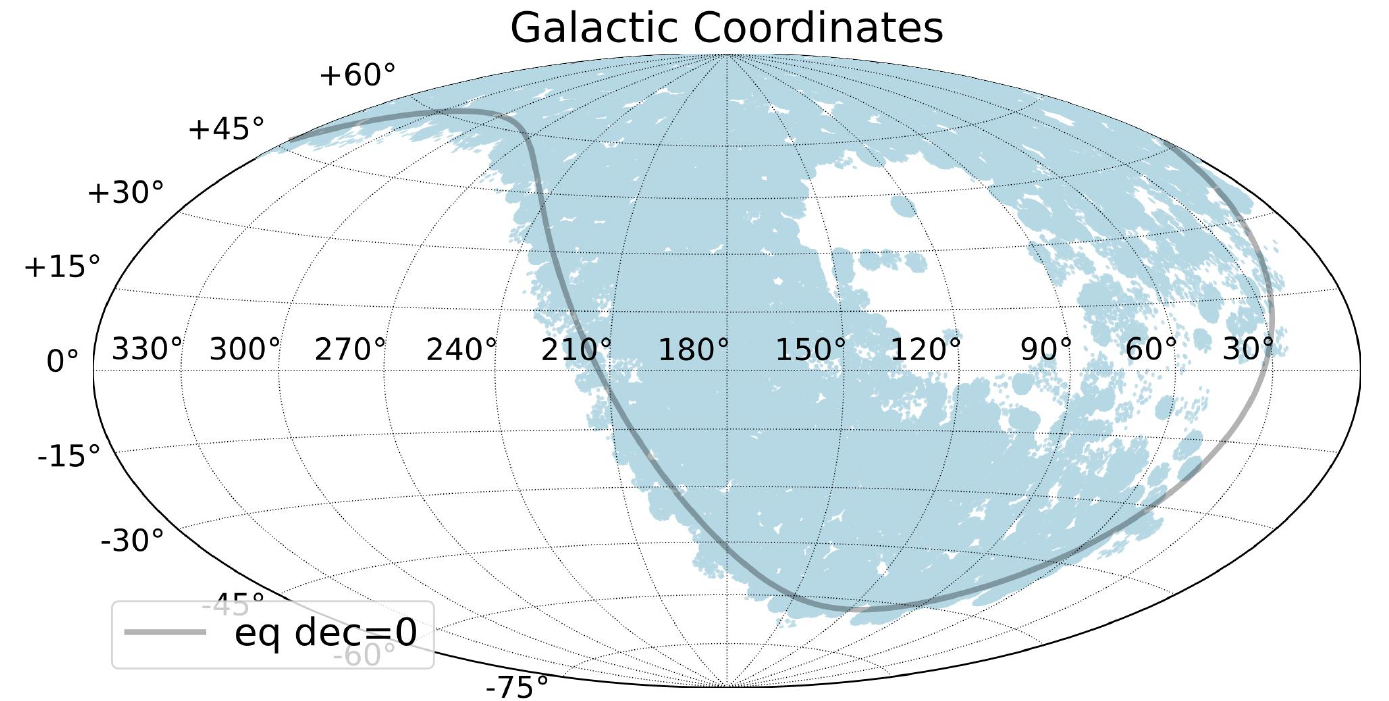}
\caption{Galactic coordinate distribution of LAMOST DR10 M dwarf stars on the celestial sphere. \label{fig:coordinates}}
\end{figure*}

However, analyzing M dwarf spectra remains challenging. Their complex spectral lines make modeling difficult and introduce uncertainties. In the optical range, only half of the spectral lines have well-determined laboratory properties \citep{2014dapb.book...39K}. This problem is more serious for cool stars where molecular lines dominate \citep{2019ARA&A..57..571J}. UV and IR wavelength data are even less complete. Crowded molecular features make normalization difficult, and parameter degeneracy adds further uncertainty.

Domain adaptation methods in transfer learning have emerged as a promising solution \citep{NIPS2017_dc6a6489,Zhu_2017_ICCV}. These methods can learn from different but related domains to address the lack of labeled data. Cycle-StarNet \citep{2021ApJ...906..130O} is one such innovation in astronomy. It calibrates theoretical spectra to match observations using unsupervised learning. This approach has improved parameter measurements for both high-resolution APOGEE data \citep{2021ApJ...906..130O} and medium-resolution LAMOST MRS spectra \citep{2023ApJS..266...40W}.

Cycle-StarNet is particularly valuable for low-resolution M dwarf spectra. The gap between observed and theoretical spectra is larger for low-mass stars than for solar-type stars. In this work, we applied Cycle-StarNet to LAMOST DR10 low-resolution M dwarf spectra. We derived consistent parameters for over 500,000 M dwarf stars - the first such comprehensive analysis using full spectral fitting. We present our data and methods in Sections \ref{sec:data} and \ref{sec:method}. Section \ref{sec:result} contains our analysis and describes the value-added catalog. In Section \ref{sec:comp}, we evaluate our parameters through comparison with other datasets. Finally, Section \ref{sec:con} summarizes our findings.

\section{Data} \label{sec:data}

\subsection{M dwarf stars from LAMOST Data Release 10} \label{subsec:dr10M}
LAMOST, located at Xinglong Observatory in Hebei, China, is a 4-meter reflecting Schmidt telescope equipped with active optics. Its multi-object fiber spectroscopic system enables simultaneous observations of 4,000 celestial targets~\citep{2012RAA....12.1197C}. As of June 2022, LAMOST has published the 10th data release with a total of over 22.29 million spectra, consisting of 11.81 million low-resolution spectra and 10.48 million medium-resolution spectra\footnote{https://www.lamost.org/dr10/v1.0/}. The low-resolution survey covers a broad spectral range of 3700-9000 \AA \ with a resolution R$\sim$1800 \citep{2015RAA....15.1095L}. In this dataset, 818,686 low-resolution spectra are classified as M-type dwarfs.

\begin{figure}[htp!]
\includegraphics[width=83mm]{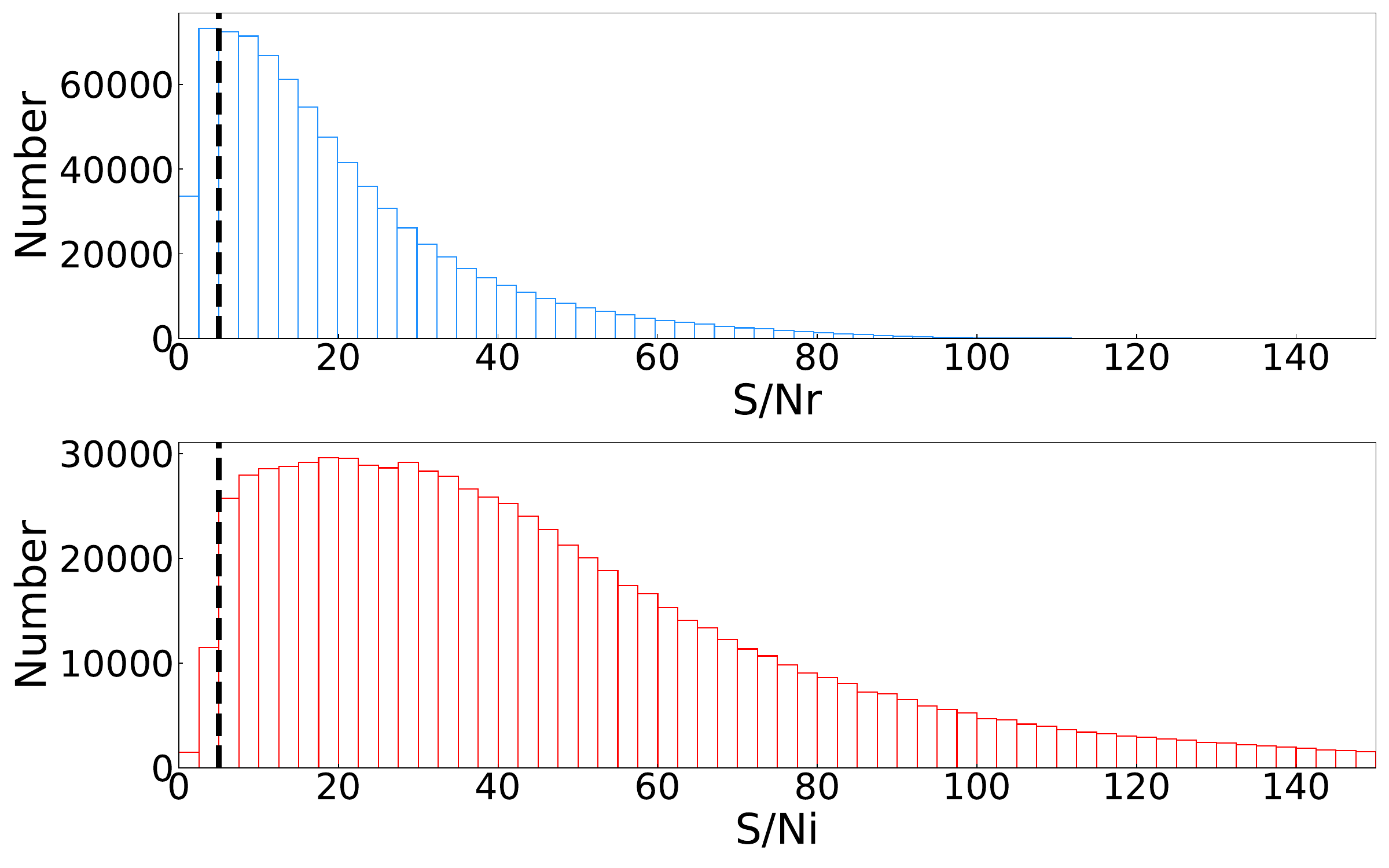}
\caption{Signal-to-noise ratio (S/N) distributions of M dwarf spectra classified by the LAMOST pipeline. The upper and lower panels show the distributions for the $r$-band and $i$-band, respectively, with black dashed lines indicating the sample selection thresholds of S/Nr = 5 and S/Ni = 5.
\label{fig:sn}}
\end{figure}

In this study, we applied the following filtering criteria to the original dataset:
\begin{enumerate}
\item We initially enforced a minimum spectral signal-to-noise ratio threshold of 5 in both r band (SNRr) and i band (SNRi), because full-spectrum fitting is employed to derive parameters from LAMOST low-resolution spectra covering both r and i bands. Figure \ref{fig:sn} shows the average spectral S/N distributions in the two bands for the original dataset.

\item We removed objects with Gaia DR3 Renormalized Unit Weight Error (RUWE) $>$1.4 to eliminate potential binary contamination. Gaia's RUWE is a normalized astrometric noise indicator that equals $\sim$1 for single stars, independent of their color or magnitude \citep{2018A&A...616A...2L}. While RUWE$>$1.4 strongly indicates binarity, it is sensitive to systems with separations of 100-1000 mas and magnitude differences $\Delta G<$3 \citep{2021AJ....162..128W}. Therefore, while our criterion helps remove many binary systems, it should be noted that some binaries, especially those outside this optimal detection range, may remain in our sample (see Section 4.5 for further discussion of binarity effects).

\item We further excluded possible giant stars and other stellar contaminants by implementing magnitude and color constraints of 4$<$$M_G$$<$15 and 1$<$BP$-$RP$<$5.

\item To identify and remove low-quality spectra, we implemented an iterative K-means clustering process. This method groups spectra with similar anomalous patterns (often resulting from systematic data reduction issues) and allows for batch inspection and removal. The process continues until all remaining clusters show typical M dwarf spectral characteristics.
\end{enumerate}
Ultimately, a total of 538,107 candidate spectra remained available for measurement.

\subsection{PHOENIX Atmospheric Model and Spectral Emulator}\label{subsec:emu}

Physically modeling the spectra of cool stars is a complex and time-consuming process. In this work, we adopted The Payne~\citep{2019ApJ...879...69T} as the spectral interpolator, which enables users to efficiently generate synthetic spectra for any specified parameter combination, which is a critical aspect for constructing training sets within the synthetic domain for domain adaptation networks.

The Payne is a multilayer perceptron neural network, designed for precise and accurate interpolation and prediction of stellar spectra across a high-dimensional label space, utilizing a moderate number of base models. These models ensure consistency in atmospheric structure and radiative transport reflective of stellar labels, thereby avoiding the need for recalibration. Consequently, even if the domain transfer process of the spectrum is unsupervised, its results can still be associated with stellar labels.

\begin{figure*}[htp!]
\includegraphics[width=170mm]{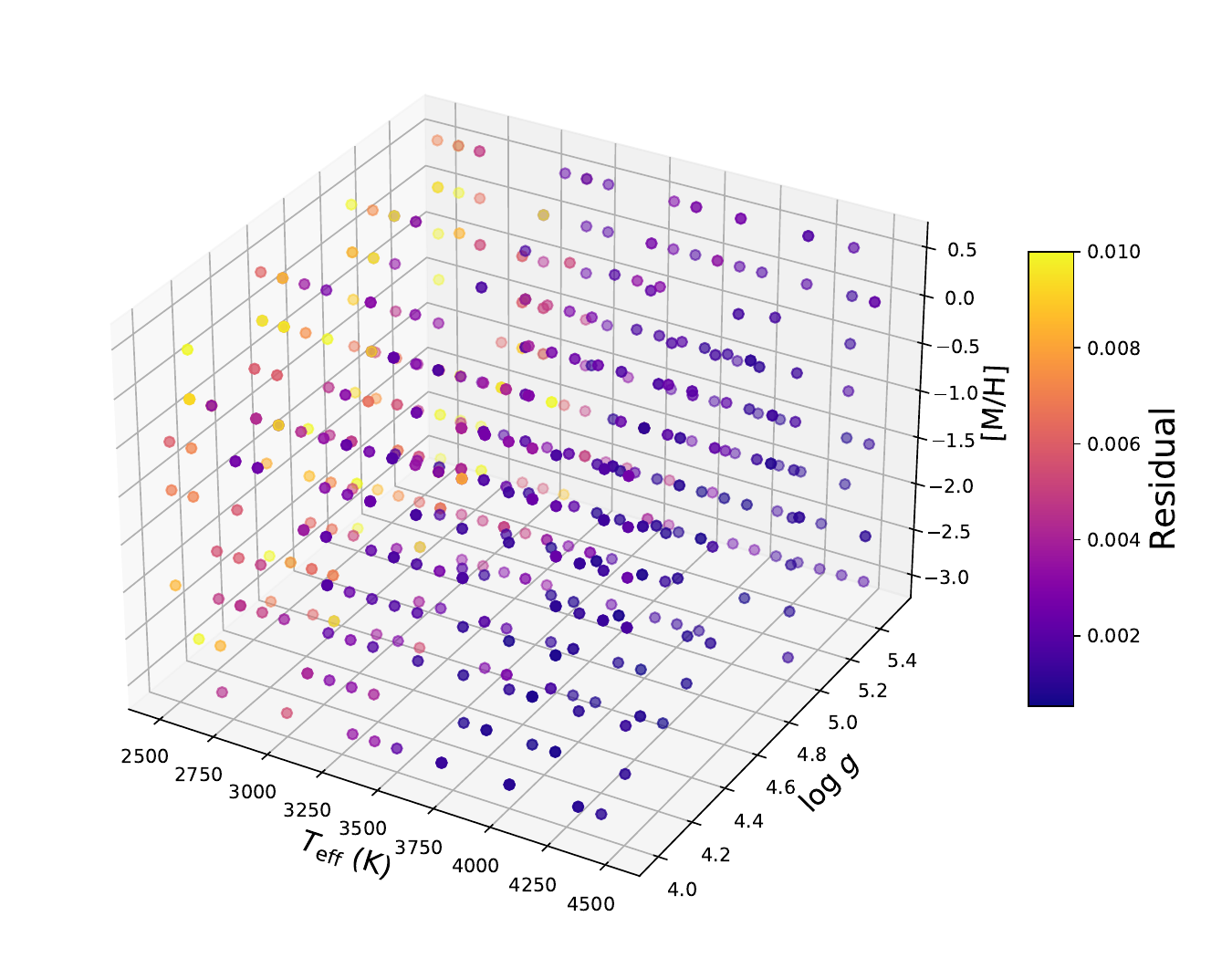}
\caption{Evaluation of the reconstruction accuracy of the spectral emulator across the entire parameter space. Each point represents a synthetic spectrum from the model grid in the validation set, with points color-coded by the reconstruction residuals, as calculated and described in Sec.~\ref{subsec:emu}. 
\label{fig:payne}}
\end{figure*}

This study utilizes the latest pre-calculated BT-Settl CIFIST model grid as the training set for Payne. The model atmospheres \citep{2012RSPTA.370.2765A,2013MSAIS..24..128A,2014ASInC..11...33A,2015A&A...577A..42B}, are computed with PHOENIX version 15.5 \citep{1997ApJ...483..390H,2001ApJ...556..357A}. These models incorporate specialized treatments for cool atmospheres (T$<$3000 K), including the Settl cloud formation model, \cite{2011SoPh..268..255C} solar abundances, and radiation hydrodynamic simulations for M-L-T dwarf atmospheres \citep{2010A&A...513A..19F, 2012JCoPh.231..919F}. The latest models include updated molecular line lists for water vapor  \citep{2006MNRAS.368.1087B}, metal hydrides (CaH, FeH, CrH, TiH; \citealt{2006AIPC..855..143B}), metal oxides (VO, TiO; \citealt{1998A&A...337..495P}), and CO2 \citep{2004SPIE.5311..102T}.

The BT-Settl synthetic spectra have been extensively validated through observational comparisons and atmospheric parameter determinations (e.g., \citealt{2013A&A...556A..15R,2014A&A...564A..90R,2015ApJ...804...64M,2017MNRAS.464.3040Z,2017ApJ...851...26V,2018A&A...620A.180R,2018A&A...610A..19R,2021ApJ...908..131Z,2021AJ....161..172D,2020AJ....159...30H, 2022ApJ...927..122H}). These studies demonstrate the models' capability to reproduce optical-NIR spectral profiles of M and L dwarfs and subdwarfs.

For this analysis, we employ a newly computed subgrid (priv. comm. Derek Homeier) previously utilized in~\cite{2020AJ....159...30H,2022ApJ...927..122H} and \cite{2023ApJ...942...40Z}. This subgrid extends beyond the classical CIFIST models in its treatment of $\alpha$-element enhancement. The parameter space encompasses $T_{\text{eff}}$ of 2500-4000 K in 100 K steps, $\log g$ of 4.5-5.5 dex in 0.5 dex increments, and [M/H] from $-$3.0 to +0.5 dex, also in 0.5 dex steps. The [$\alpha$/Fe] values, varying in 0.2 dex increments, are metallicity-dependent: ranging from $-$0.2 to +0.2 dex for [M/H] $\geq$ 0.0, 0.0 to +0.4 dex for [M/H] = $-$0.5, and +0.2 to +0.6 dex for [M/H] $\leq$ $-$1.0.

In the training process, we convolved the synthetic spectra to a resolution of R$\sim$2000 and randomly allocated all synthetic spectra within the model grid among the training and validation sets in an 8:2 ratio. Every synthetic spectrum is convoluted to the LAMOST resolution and then normalized and masked as described in section \ref{subsec:process}. 

Figure \ref{fig:payne} evaluates the precision of reconstruction of the validation set, the residuals are calculated as follows

\begin{equation}
    \frac{1}{N} \sum_{i=1}^N \frac{\left| F_{\text{model}, i} - F_{\text{recon}, i}\right|}{F_{\text{model}, i}} ,
\end{equation}
\noindent
where $N$ represents the number of observation points, and $F_{\text{model}, i}$ and $F_{\text{recon}, i}$ denote the model and emulated values at the $i$-th observation point, respectively. The results show that The Payne can achieve a spectral reconstruction accuracy of higher than 1\%. In the parameter space where most of our observational dataset lies, we achieve better reconstruction accuracies of less than 0.5\%. Once trained, it can generate a synthetic spectrum with a given parameter combination in one CPU second, contrasting with the time-consuming nature of direct spectrum synthesis.

\subsection{Radial Velocity Measurement}
The first step in processing observed spectra is to measure radial velocities (RVs) and shift the spectra to their rest-frame wavelengths. In this study, RV of each observed spectrum was derived using the fitting module of The Payne from the Doppler shift. These measurements were compared to radial velocities provided in Gaia Data Release 3 (DR3) for common sources. The Gaia Radial Velocity Spectrometer (Gaia-RVS; \citealt{2018A&A...616A...5C}) is an integral-field spectrograph with a resolving power of 11,500, covering the infrared wavelength range of 8450-8720 $\rm \AA \ $. The second release of Gaia DR3 provided precise RV measurements for over 33 million stars \citep{2023A&A...674A...5K}.

\begin{figure}[htp!]
\centering
\includegraphics[width=87mm]{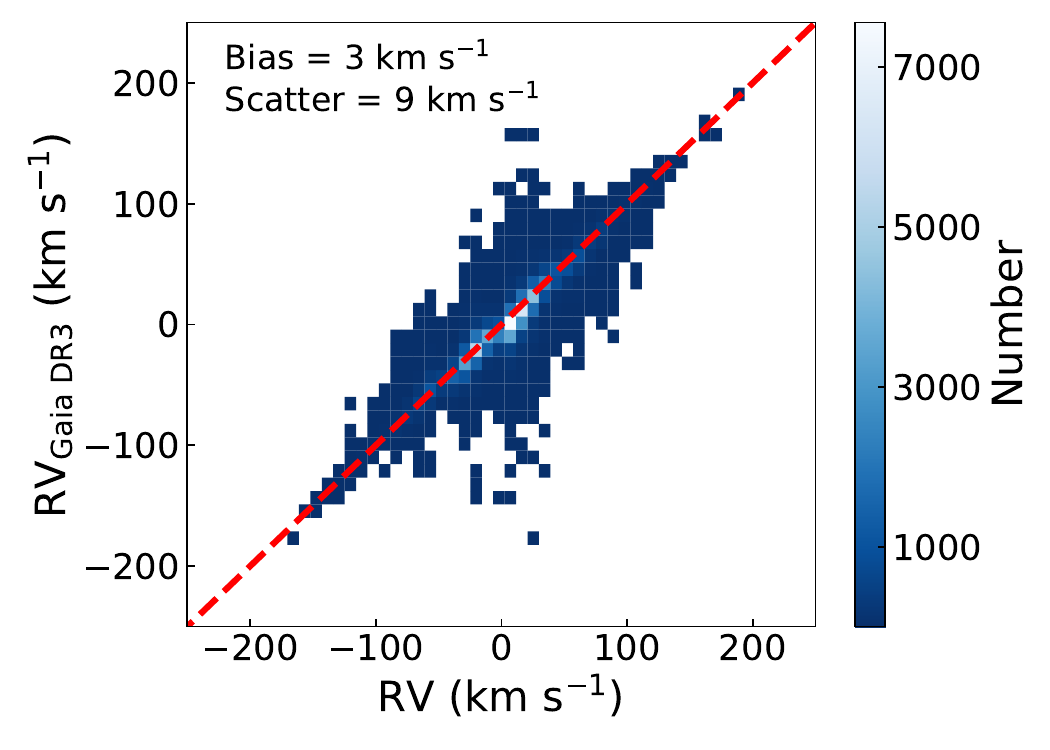}
\caption{Comparison between Gaia RV and RV measurements from this study for approximately 100,000 homologous samples. The density map is color-coded by the sample count. The red dashed line marks the reference line.
\label{fig:RV_gaia}}
\end{figure}

We cross-matched our catalog with the Gaia DR3 dataset and identified 100,435 common sources for comparison. The comparative analysis is presented in Figure \ref{fig:RV_gaia}. The result demonstrates that our radial velocity measurements are in good agreement with those from Gaia, showing a bias of 3 km s$^{-1}$ and a dispersion of 9 km s$^{-1}$. This confirms a greater accuracy of our measurements compared to the results provided by LAMOST Stellar Parameter pipeline (LASP; \citealt{2011RAA....11..924W,2014IAUS..306..340W,2015RAA....15.1095L}): in a similar analysis, LASP's RVs exhibited a bias of $-$5 km s$^{-1}$ and a dispersion of 11 km s$^{-1}$ with comparison to Gaia DR3 results.

The errors of radial velocities are calculated from two components: the fitting error and the measurement error induced by spectral noise, which varies with the S/N. We estimated how measurement errors change with spectral quality using 29,238 sources that had multiple observations (n$\geq$3). The specific process was as follows: (1) For all sources with multiple spectral observations, we calculated the average radial velocity $\bar{\text{RV}}$ of each source from their spectra observed under all conditions, and then the difference between the measured radial velocity of an individual spectrum and this average value, i.e., $\Delta$RV, was set as the error of the measurement for the spectrum. (2) We grouped all the $\Delta$RVs from 29,238 sources into multiple bins based on S/N and then fitted the relationship between the 1$\sigma$ values of $\Delta$RVs in each bin and the corresponding S/N. Finally, using this empirical relationship, we determined the S/N-dependent RV error for each spectrum based on its signal-to-noise ratio. 

The process is similar to the determination of precision of other parameters ($T_{\rm eff}$, $\log g$, [M/H]) described in Section \ref{subsec:params}. The fitted empirical relationship is shown in Figure \ref{fig:prec_func}.

\begin{figure}[htp!]
\centering
\includegraphics[width=90mm]{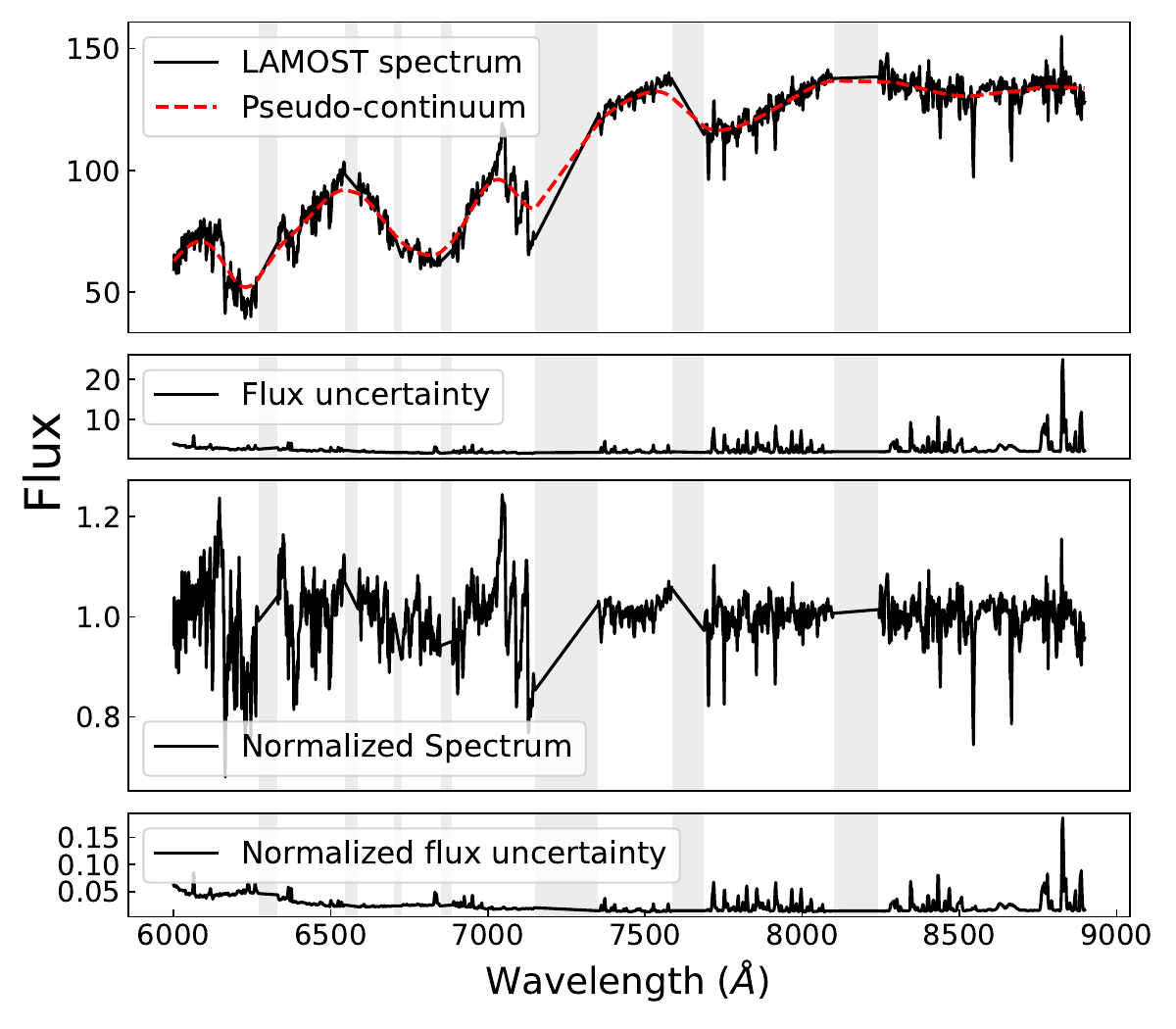}
\caption{Example of a LAMOST observed spectrum and its processing. From top to bottom: The original spectrum (black) with its pseudo-continuum fit (red); the flux uncertainty; the normalized spectrum after continuum division; and the normalized flux uncertainty. Gray shaded regions indicate masked wavelength ranges that are excluded from analysis. The wavelength range covers from 6000 to 8900 $\text{\AA}$.
\label{fig:spectrum}}
\end{figure}

\subsection{Spectral Pre-processing}\label{subsec:process}

The normalization of M-type stellar spectra presents challenges due to the dense molecular line forests, which obscure the continuum spectrum at low resolutions. This complexity often results in what can only be termed a ``pseudo-continuum'' even in regions of minimal opacity. In this study, since normalization was only intended to produce spectral data suitable for training models, and the subsequent domain adaptation module can correct errors introduced during the normalization process, each spectrum was normalized by dividing by the corresponding smoothed flux F($\lambda$) with F($\lambda$) defined as 

\begin{equation}
    F(\lambda) = \frac{\sum_{i}{f_iw_i(\lambda)}}{\sum_i(w_i(\lambda))},
\end{equation}
\noindent
where $f_i$ is the flux at $\lambda_i$ and the weight $w_i(\lambda)$ is a Gaussian function

\begin{equation}
    w_i(\lambda)=e^{-\frac{(\lambda-\lambda_i)^2}{\sigma^2}}.
\end{equation}

$\sigma$ is set to be 30 \AA \ \ to convolve the entire spectrum. Additionally, we applied the same normalization process to the errors associated with each observed spectrum, using the smoothed spectrum as the normalization reference. In the training process, the ``normalized'' errors for synthetic spectra are fixed to be 0.01.

For our spectroscopic analysis, we adopted spectral data in the wavelength range of 6000-8900 \AA, which encompasses most prominent spectral features of M dwarf stars. The spectra were resampled at a consistent interval of 1 \AA \ \ into a uniform array. Additionally, specific wavelength ranges were masked during the spectral fitting process to mitigate potential contamination from telluric absorption or chromospheric activity: 6270-6330 \AA, 6546-6586 \AA, 6700-6725 \AA, 6850-6885 \AA, 7150-7350 \AA, 7587-7687 \AA, and 8100-8240 \AA. Figure \ref{fig:spectrum} shows an example of our spectral processing procedure applied to a typical LAMOST spectrum.

\section{Method} \label{sec:method}

Here, we introduce a deep-learning domain adaptation method Cycle-StarNet \citep{2021ApJ...906..130O}, which can bridge the gap between theoretical and observed spectra by learning hidden aspects of the two domains automatically. The domain of synthetic data here is referred to as the ``synthetic domain'' and the observed data as the ``observed domain''. 

\subsection{Domain Adaptation Network}
The domain adaptation network minimizes the synthetic gap, addressing the discrepancies caused by imperfections in the stellar atmospheric model, observational effects, and data processing errors. Referring to Cycle-StarNet, this goal is achieved by two primary components:

(1) A hierarchical structure with multiple encoder-decoder pairs operating at different levels: two low-level pairs handle domain-specific features separately for synthetic and observed spectra, while a high-level pairs abstract shared physical characteristics across domains. The encoders progressively compress data into lower-dimensional representations, with decoders reconstructing the original form. This design efficiently separates domain-specific learning at lower levels from the abstraction of shared physical concepts at higher levels, with an additional split encoder-decoder for observed data improving convergence. This process also implicitly denoises the spectra, making it possible to work directly with noisy training spectra.

(2) A generative adversarial network (GAN; \citealt{2014arXiv1406.2661G}) ensures that the abstractions in the latent spaces of the two domains are shared. To accomplish this, the spectra in the observed domain are dynamically mapped to a ``shared latent space'' common to the synthetic domain. This is enforced to ensure uniformity, along with a separate ``split latent space'', which accounts for discrepancies attributed to the synthetic gap. The physical meaning is preserved throughout the transfer by introducing a cycle-consistency constraint, whereas adversarial learning is applied to the latent space to appropriately allocate shared and split information.

Note that this process is unsupervised; it does not require the synthetic and observed spectral datasets to have a one-to-one correspondence, nor do they need to have the same stellar labels. However, it operates on the basis that their physical parameter spaces overlap.

Once well-trained, the network learns to link two sets of unlabeled spectra and morph between domains. In other words, it can transfer spectral models from the synthetic to the observed domain, or vice versa, correcting systematic discrepancies between the two domains.

\subsection{Training Process and Loss Functions}

The data from the two domains are spectral fluxes of the training sets and denoted as $\mathcal{X}_{synth}$ and $\mathcal{X}_{obs}$, respectively. The training process iteratively adjusts the network's parameter weights by minimizing a loss function through gradient descent. Following the core idea of the domain adaptation network, the model's loss function includes components from two auto-encoders, two cross-domain generations, and the discriminator.

The training consists of three parts:

(1) Ensuring encoding-decoding pairs complete within-domain reconstruction, i.e., maximizing the consistency between $\mathcal{X}_{synth}$ and $\mathcal{X}_{synth\to synth}$, $\mathcal{X}_{obs}$ and $\mathcal{X}_{obs\to obs}$. 

The loss function of within-domain reconstruction is denoted as $\mathcal{L}_{rec}$,
\begin{equation}
\begin{split}
    \mathcal{L}_{rec,synth} &= d(\mathcal{X}_{synth}, \mathcal{X}_{synth\to synth})\\
    \mathcal{L}_{rec,obs} &= d(\mathcal{X}_{obs}, \mathcal{X}_{obs\to obs}),
\end{split}
\end{equation}
\noindent
where $d$ is the distance function, specifically a standard mean squared error (MSE) with samples weighted by the spectrum uncertainties.

(2) Applying generative adversarial networks to achieve cross-domain transfer, i.e., maximizing the consistency between $\mathcal{X}_{obs}$ and $\mathcal{X}_{synth\to obs}$, $\mathcal{X}_{synth}$ and $\mathcal{X}_{obs\to synth}$. This network includes generators and discriminators, which train simultaneously but with different objectives: generators are responsible for producing counterfeit samples as realistic as possible, while discriminators are responsible for identifying whether given samples are counterfeit. 

In this work, the cross-domain auto-encoders play the role of generators, while the discriminators are responsible for accepting a cross-domain transferred spectrum and a within-domain reconstructed spectrum, and making correct judgments on them. In addition to discriminating between the spectral pairs, the discriminators evaluate the latent space representations that generate these spectra, providing additional constraints on the model. The training objective is implemented using the binary cross-entropy function. The loss function is denoted as $\mathcal{L}_{adv}$,
\begin{equation}
\begin{split}
    \mathcal{L}_{adv,synth} &= H(1, C_{synth}(\mathcal{X}_{synth\to synth}, \mathcal{Z}_{sh}))\\
                 &+ H(0, C_{synth}(\mathcal{X}_{obs\to synth}, \mathcal{Z}_{sh}))\\
    \mathcal{L}_{adv,obs} &= H(1, C_{obs}(\mathcal{X}_{obs\to obs}, (\mathcal{Z}_{sh}, \mathcal{Z}_{sp})))\\
                &+ H(0, C_{obs}(\mathcal{X}_{synth\to obs}, (\mathcal{Z}_{sh}, \mathcal{Z}_{sp}))),
\end{split}
\end{equation}
\noindent
where $H$ represents the cross-entropy function, $C_{synth}$ and $C_{obs}$ represent the discriminators for the theoretical and observational domains, respectively, and ($\mathcal{Z}_{sh}$, $\mathcal{Z}_{sp}$) represents the expression of the observed domain spectrum generated jointly through the shared latent space and split latent space. When both the input spectrum and latent space expression are judged as reconstructed, i.e., real, the discriminator outputs 1, and 0 for ``fake'' cross-domain transferred objects. When the discriminators of each domain can correctly identify reconstructed spectra and cross-domain transferred spectra, $\mathcal{L}_{adv}$ takes a value of 0. Therefore, the objective of the discriminator is to minimize this loss function, while the objective of the generative network is the opposite, to maximize it, thereby achieving the adversarial goal and realizing mutual improvement of the two networks during the training process.

(3) Cycle reconstruction, i.e., $\mathcal{X}_{synth\to obs\to synth}$ and $\mathcal{X}_{obs\to synth\to obs}$. This is a cycle-consistency constraint introduced to ensure that the cross-domain generated spectrum for a given spectrum is the correct corresponding spectrum in the opposite domain. The loss function is denoted as $\mathcal{L}_{cr}$,

\begin{equation}
\begin{split}
    \mathcal{L}_{cr,synth} &= d(\mathcal{X}_{synth}, \mathcal{X}_{synth\to obs\to synth})\\
    \mathcal{L}_{cr,obs} &= d(\mathcal{X}_{obs}, \mathcal{X}_{obs\to syn\to obs}),
\end{split}
\end{equation}
\noindent
where the definition of the distance function $d$ is the same as in $\mathcal{L}_{rec}$.

Finally, the total loss is represented as
\begin{equation}
\begin{split}
    \mathcal{L}&= \lambda_1(\mathcal{L}_{rec, synth} + \mathcal{L}_{cr, synth})\\
     &+ \lambda_2(\mathcal{L}_{rec, obs} + \mathcal{L}_{cr, obs})\\
     &+ \lambda_3(\mathcal{L}_{adv, synth} + \mathcal{L}_{adv, obs}),
\end{split}
\end{equation}
\noindent
where $\lambda_1$, $\lambda_2$, and $\lambda_3$ are hyper-parameters used to control the weight of each part.

\begin{figure*}[htp!]
\centering
\includegraphics[width=175mm]{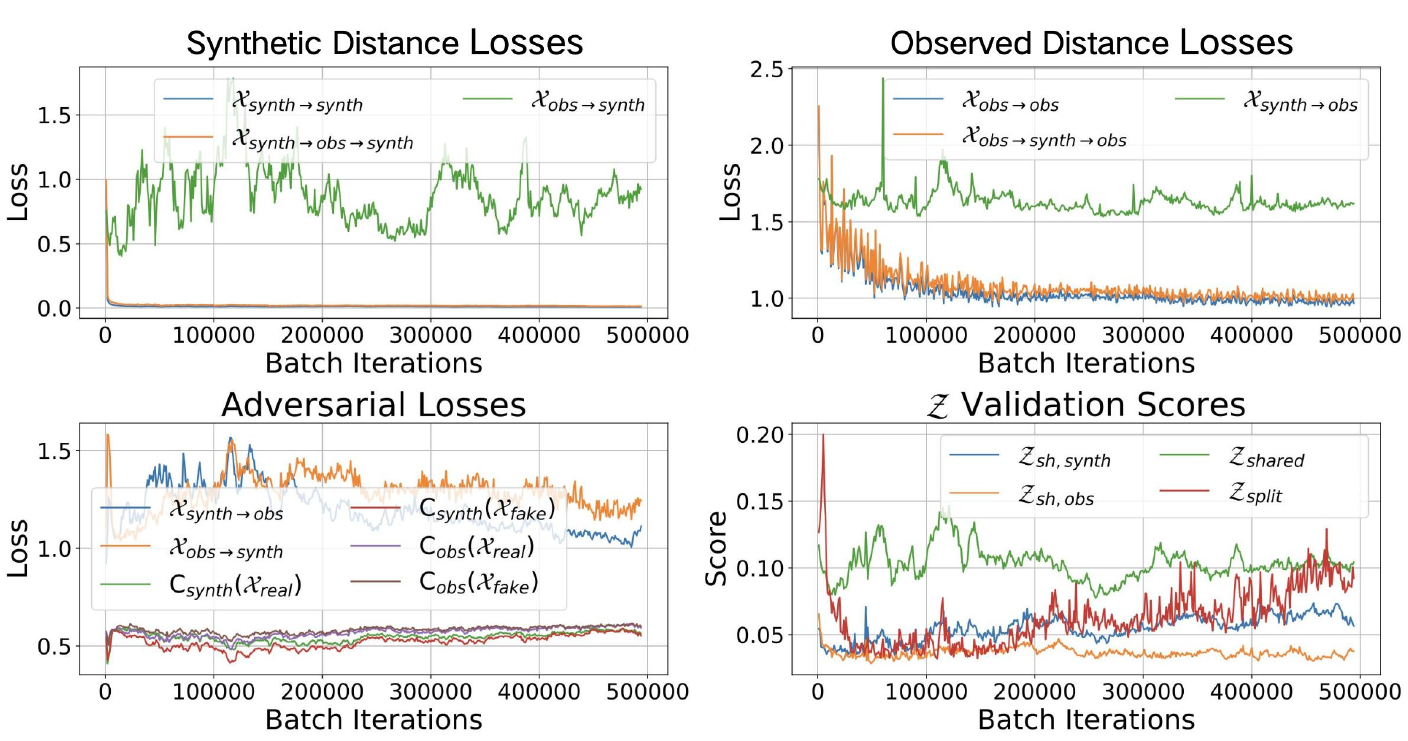}
\caption{Training losses of Cycle-StarNet as a function of epochs. The upper panel displays the losses in both the synthetic and observed domains for the within domain reconstructions ($\mathcal{X}_{\mathrm{synth \rightarrow synth}}$, $\mathcal{X}_{\mathrm{obs \rightarrow obs}}$), cross-domain transferring ($\mathcal{X}_{\mathrm{synth \rightarrow obs}}$, $\mathcal{X}_{\mathrm{obs \rightarrow synth}}$), and cycle reconstructions ($\mathcal{X}_{\mathrm{synth \rightarrow obs \rightarrow synth}}$, $\mathcal{X}_{\mathrm{obs \rightarrow synth \rightarrow obs}}$). The left lower panel shows the evaluation of the adversarial losses on the training set for both the transfer loss ($\mathcal{X}_{\mathrm{synth \rightarrow obs}}$, $\mathcal{X}_{\mathrm{obs \rightarrow synth}}$) and the adversarial loss ($C_{\mathrm{synth/obs}}(\mathcal{X}_{\mathrm{real/fake}})$. The right lower panel shows the validation scores for the latent spaces ($\mathcal{Z}_{shared/split}$). \label{fig:loss_function}}
\end{figure*}

\begin{figure*}[htp!]
\centering
\includegraphics[width=165mm]{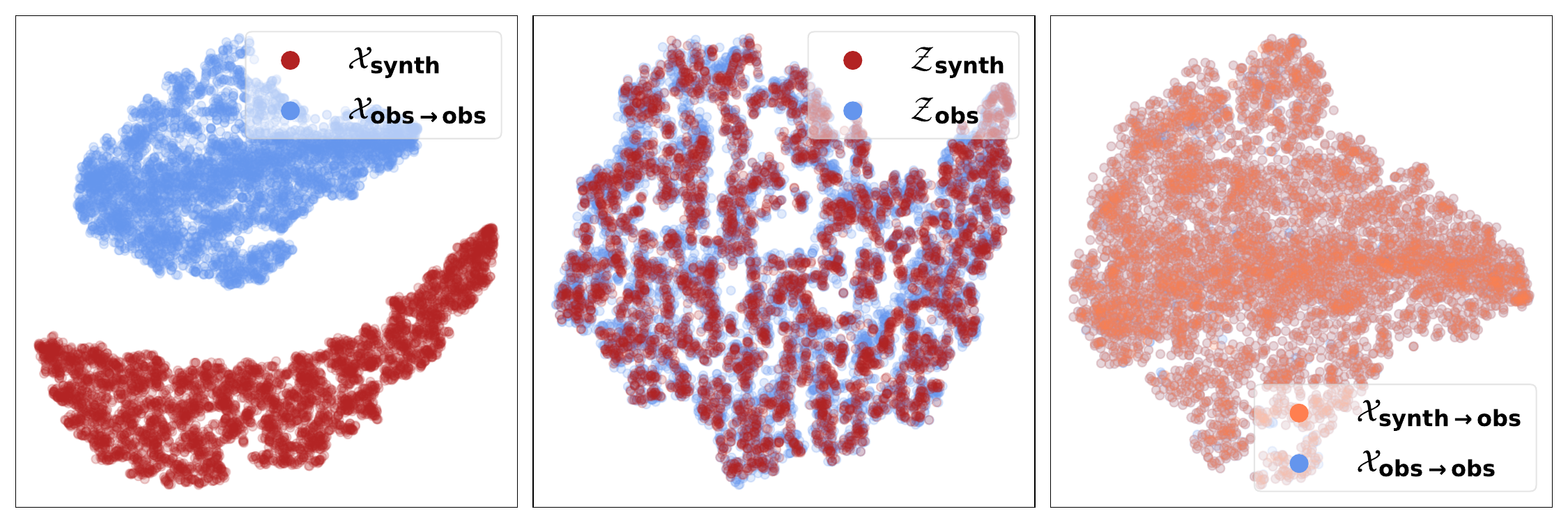}
\caption{The t-SNE projections of 10,000 pairs of test spectra. The left panel highlights the disparity, referred to as the ``synthetic gap'', between PHOENIX models and LAMOST spectra. The middle panel demonstrates that a shared representation for the latent spaces of the two domains has been achieved after model training. The right panel illustrates the transfer results in the observed domain, showcasing success in bridging the synthetic and observed domains through domain adaptation.
\label{fig:tsne}}
\end{figure*} 

\subsection{Assembly of the Training Sets}
The training data include the synthetic spectral dataset and the observed spectral dataset, each dataset serves as the input and output of its corresponding auto-encoders and also the verification input of the adversarial generative network. As described above, this is not a supervised learning method. The training sets in the two domains do not require one-to-one correspondence but only covering the same parameter space.

\subsubsection{observed domain}
We randomly selected 110,000 LAMOST low-resolution spectra from our candidate dataset (Section \ref{subsec:dr10M}) as the observed domain training set, with an 8:2:1 split for training, validation, and testing. 

Data-driven methods commonly rely on high-quality training spectra, potentially limiting their performance on low S/N data. Given our approach's ability to separate physical components from noise, we chose to train without S/N constraints.

\subsubsection{Synthetic domain}\label{subsubsec:syn_train}

\begin{figure*}[htp!]
\centering
\includegraphics[width=165mm]{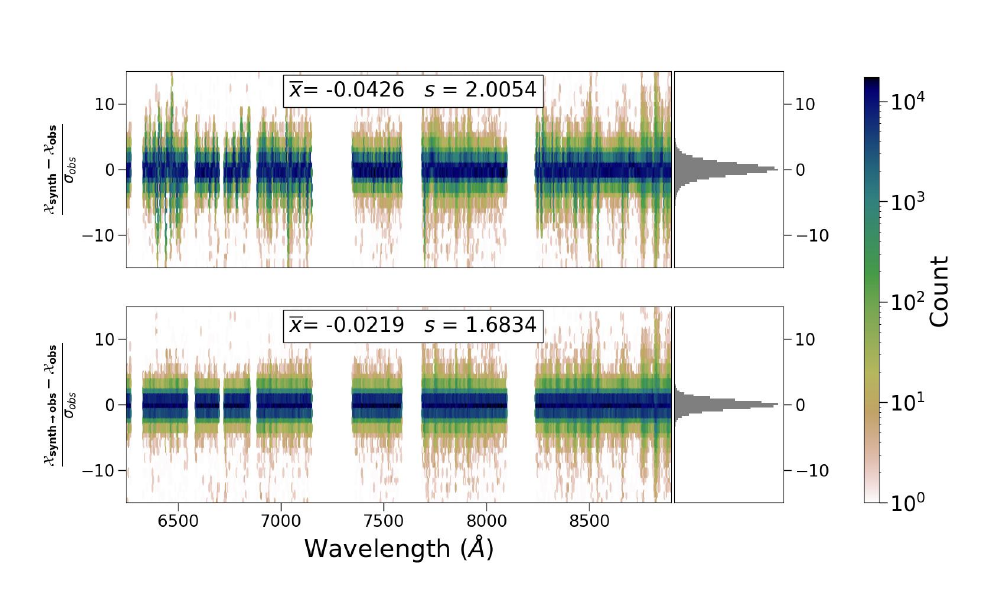}
\caption{Comparison of residuals for 10,000 LAMOST spectra in the test set, showing the residuals between the best-fit and PHOENIX spectra both before and after transfer. After applying the domain adaptation model, the spectral residuals decreased from the mean intrinsic spectral error shown in the upper panel (over 2 times) to 1.68 times, as shown in the lower panel, with the fitting accuracy improved by around 20\%.
\label{fig:fitting_residuals}}
\end{figure*}

To construct the training dataset for the synthetic domain, we first built a set of parameter combinations covering a similar parameter space as the observed domain training set, then generated corresponding synthetic spectra based on these parameter combinations via The Payne. The procedure includes four steps.

(1) We utilized theoretical isochrones generated from the PAdova and TRieste Stellar Evolution Code (PARSEC;~\citealt{2012MNRAS.427..127B}) with specified criteria (3000 K$\leq T_{\rm eff}\leq$4000 K, 4.0 dex$\leq
\log g\leq$6.0 dex, $-$1.0 dex$\leq$[M/H]$\leq$0.5 dex, and Age=5 Gyr) as the foundational dataset to derive initial parameter combinations in the ($T_{\rm eff}$, $\log g$, [M/H]) three-dimensional parameter space. The [$\alpha$/M] values were set empirically as 0.0 for [M/H]$\geq$0, 0.2 for [M/H]=$-$0.5 dex, and 0.4 for [M/H]=$-$1.0 dex.

(2) Based on these initial discrete parameter combinations, we applied a Gaussian mixture model to generate over one million continuous parameter combinations distributed across the four-dimensional parameter space ($T_{\rm eff}$, $\log g$, [M/H], [$\alpha$/M]).

This parameter generation approach ensures physically meaningful $\log g$ values for each combination of temperature and metallicity while covering the parameter space of LAMOST M dwarf samples. For low-mass stars like M dwarfs, whose initial conditions are primarily determined by mass and metallicity, evolution becomes extremely slow after entering the main sequence phase, with their position on the Kiel diagram remaining nearly constant. We chose 5 Gyr isochrones as the baseline because the $T_{\rm eff}$--$\log g$ relations agree within 0.03 dex in $\log g$ for ages between 1 and 10 Gyr in the temperature range of M dwarf stars up to 4000 K. In practice, this method of determining $\log g$ has been adopted in spectroscopic parameter measurements to break parameter degeneracy, as demonstrated in \cite{2018A&A...615A...6P,2019A&A...627A.161P}. In their approach, the temperatures and other parameters of M dwarf stars were first measured by fitting high-resolution spectra with a model grid, then the $\log g$ values were calculated based on the $T_{\rm eff}$--$\log g$ relations from 5 Gyr evolutionary models. It should be noted that this method is less accurate for stellar objects with age estimates younger than 1 Gyr; thus, we removed these objects from our final catalog, as detailed in Section \ref{sec:result}.

At this step, while our generated parameter combinations covered the observed parameter space, their distribution differed from that of the observed domain. To ensure that the distribution closely matched that of the observed training sample and to create a final training set that both reflected the characteristic distribution of the LAMOST dataset and avoided systematic bias from any single source dataset, we performed a secondary sampling in the dimensions of $T_{\rm eff}$ and [M/H] in the following third step. The reference distribution was derived from a combined distribution of LAMOST-common samples from multiple publications.

(3) We achieved further refinement by executing co-distributed sampling on the temperature and metallicity components of the ensemble derived above. This approach enabled us to apply precise constraints on the parameter space.

To establish the reference distribution, we selected samples from multiple publications/survey datasets that share common sources with LAMOST spectra, combining their temperature and metallicity measurements. We compiled the samples from LAMOST DR10 Low Resolution Catalog cross-matched with data sets from \cite{2012ApJ...748...93R,2014A&A...564A..90R,2015ApJ...804...64M,2017ApJ...836...77Y,2018A&A...615A...6P,2018A&A...620A.180R,2019AJ....157...63K,2019ApJ...878..134K,2019A&A...627A.161P,2020ApJ...892...31B,2020AJ....159...30H,2020A&A...642A..22P,2021A&A...649A.147S} and APOGEE DR16 (calibrated). We retained stars classified as ``dM'' by LASP and those later than K5 with effective temperatures in the literature below 4000 K. The final sample exceeded 10,000 objects.

The top three contributing sources were the following.
\begin{enumerate}
\item \cite{2020ApJ...892...31B}: 5288 objects. The parameters were derived from the APOGEE spectra using The Cannon~\citep{2015ApJ...808...16N}, trained on the parameters from \cite{2015ApJ...804...64M}. \cite{2015ApJ...804...64M} determined temperatures through an empirically calibrated spectroscopic method using optical spectra, validated against interferometric measurements where available, while metallicities were calibrated using FGK binaries.
    
\item APOGEE DR16 (calibrated): 4571 objects. The SDSS collaboration adjusted APOGEE DR16 $T_{\rm eff}$ values to match the photometric scale of \cite{2009A&A...497..497G}, using linear relations as functions of metallicity and spectroscopic effective temperature (see \citealt{2020AJ....160..120J}), based on fitting to the MARCS grid.

\item \cite{2021A&A...649A.147S}: 104 objects. The parameters were derived by comparing APOGEE observations against synthetic spectra created with iSpec, Turbospectrum, and MARCS model atmospheres, using a line list containing over a million water lines.
\end{enumerate}

(4) Finally, we used this refined ensemble of parameter combinations to produce synthetic spectra using The Payne framework, establishing the training dataset for the synthetic domain. The resulting spectra were then partitioned into training, validation, and test sets, maintaining the same quantities and ratio as the observed domain to ensure robust model training and evaluation.

\subsection{Training and Testing}

The model was trained for a total of 500,000 epochs, taking about 10 GPU hours using a piece of NVIDIA Tesla V100. Figure \ref{fig:loss_function} shows the change of training loss and critics during the training process with epochs. We used 10,000 test samples to test the domain adaptation performance based on the t-SNE algorithm~\citep{JMLR:v9:vandermaaten08a} which can keep the relative distances of the samples with high dimensional data after mapping into 2-dimensional space. The trained network reduced the gap between the synthetic spectra and the observed spectra which was originally quite different, as shown in Figure \ref{fig:tsne}. Also, the fitting residuals of the observed spectra compared to their best fitting synthetic spectra decreased from over 2 times of the flux uncertainties to 1.68 times, as shown in Figure ~\ref{fig:fitting_residuals}.

\subsection{Label Derivation}

After training the spectral emulator and domain adaptation network models, we can fit the observed spectra to obtain stellar parameters. There are two options:

\begin{enumerate}

\item Fitting in the synthetic domain. The observed spectrum is transferred to the synthetic domain through Cycle-StarNet and fitted by The Payne as:
\[
\hat{y} = \underset{y}{\text{arg min}} \left( \text{Payne}(y) - \mathbf{CSN}_{\text{obs} \to \text{synth}} (F_{\text{obs}}) \right)^2,
\] 
where $y$ represents stellar parameters, $\mathbf{CSN}_{\text{obs} \to \text{synth}}$ is Cycle-StarNet transformer that transfers a spectrum from the observed domain to the synthetic domain, and $F_{\text{obs}}$ is observed spectral fluxes.

\item Fitting in the observed domain. Synthetic spectra generated by The Payne are transferred to the observed domain through Cycle-StarNet for fitting with the observed spectra as:
\[
\hat{y} = \underset{y}{\text{arg min}} \frac{1}{\sigma^2_\text{obs}} \left( F_{\text{obs}} - \mathbf{CSN}_{\text{synth} \to \text{obs}} (\text{Payne}(y)) \right)^2,
\] 
where $y$ is stellar parameters, $\mathbf{CSN}_{\text{synth} \to \text{obs}}$ is Cycle-StarNet transformer that transfers a spectrum from the synthetic domain to the observed domain, $F_{\text{ obs}}$ is observed spectral flux and $\sigma_\text{obs}$ is the flux uncertainty. 

\end{enumerate}

In this work, we used the parameters derived in the synthetic domain as the initial parameters, then we applied them by forward modeling into the observed domain for the final catalog release. In the second step, L-BFGS (the limited-memory Broyden-Fletcher-Goldfarb-Shanno algorithm; \citealt{lbfgs}) is applied to each objective. This approach allows inaccuracies to be accommodated by generating synthetic counterparts for the observed spectra. 

\section{Results} \label{sec:result}

Based on the trained network, we extracted the basic parameters ($T_{\rm eff}$, $\log g$, [M/H]) for each spectrum with S/N$>$5 in both r and i bands. We applied the following criteria to refine our catalog:

(1) We removed stellar objects with extreme parameters: stars with $T_{\rm eff}>$4300 K and $\log g$ near our fitting boundaries. We also excluded sources where the uncertainties in the three parameters and radial velocity exceeded their respective 5$\sigma$ thresholds.

(2) To ensure the overall quality of the spectral fitting process, we removed 99 stars with large $\chi^2$ values ($\chi^2>$50).

(3) We predicted ages for each star using Zoomies~\citep{2024arXiv240309878S} and excluded stars with median or mean age estimates below 1 Gyr to focus on more evolved populations. As noted in Section \ref{subsubsec:syn_train}, while we assumed a mean age of 5 Gyr for the synthetic training sets, stars much younger than 1 Gyr would have much different gravity values from older populations.

After applying these criteria, the final catalog contains 507,513 objects, corresponding to 426,008 unique stars. All subsequent analyses and comparisons (unless otherwise specified) are based on this catalog.

\begin{deluxetable}{lll}[htp!]
\centering
\tiny
\tablecaption{Data description of the value-added catalog. \label{tab:desrcription}}
\tablewidth{70pt}
\tablehead{
\colhead{Column} &
\colhead{Unit} & 
\colhead{Description} }
\startdata
ID	& - & Designation for the stellar object \\
obsid & - & Unique identifier for LAMOST spectrum \\
RA & degrees & Right Ascension (J2000) \\
Dec & degrees & Declination (J2000) \\
SNRr & - & Spectral signal to noise ratio in $r$ band \\
SNRi & - & Spectral signal to noise ratio in $i$ band \\
Subclass & - & Spectral subclass from LAMOST pipeline \\
SpT & - & Spectral subtype from SpT-TiO5 relation \\ 
Gmag& mag & Gaia DR3 G-band mean magnitude \\
BPmag & mag & Gaia DR3 integrated BP mean magnitude \\
RPmag & mag & Gaia DR3 integrated RP mean magnitude\\
Plx & pc & Gaia DR3 parallax\\
e\_Plx & pc & Gaia DR3 parallax error\\
RUWE & - & Gaia Renormalized Unit Weight Error\\
RV & km s$^{-1}$ & Radial velocity measured in this study \\
e\_RV & km s$^{-1}$ & Typical uncertainty of radial velocity \\
Teff & K & Estimated effective temperature \\
e\_Teff & K &  Typical uncertainty of $T_{\rm eff}$\\
Logg & dex & Estimated surface gravity  \\
e\_Logg & dex &  Typical uncertainty of $\log g$  \\
$\text{[M/H]}$ & dex & Estimated overall metallicity\\ 
e\_$\text{[M/H]}$ & dex &  Typical uncertainty of [M/H]\\
Activity & - & 0=default, 1=H$\alpha$ emission detected\\
VBFlag & - & 0=default, 1(2)=variable(binary) candidate
\enddata
\tablecomments{Description of each column in the value-added catalog provided in this study.}
\end{deluxetable}

\subsection{Atmospheric Parameters and Metallicity}\label{subsec:params}

\begin{figure}[htp!]
\centering
\includegraphics[width=90mm]{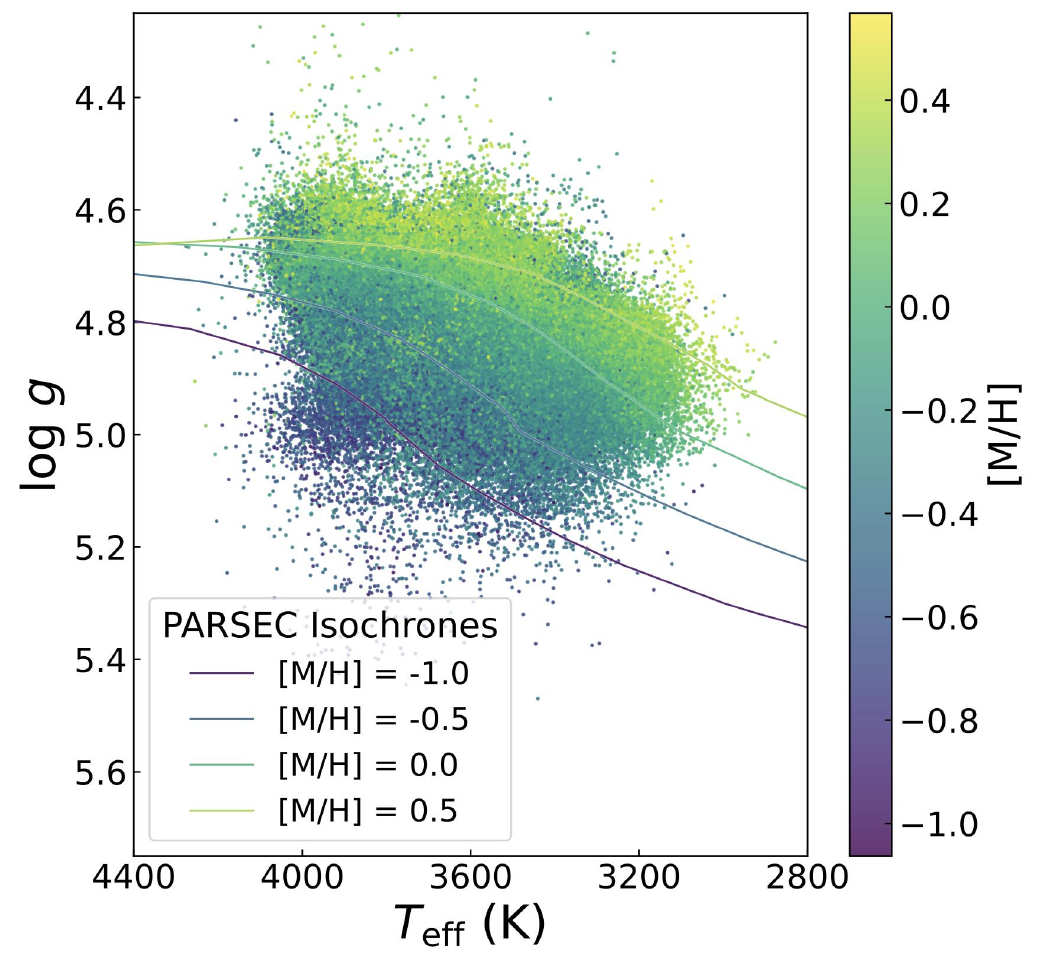}
\caption{Distribution of $\sim$500,000 M dwarf stars measured in this study on the Kiel diagram ($\log g$ vs. $T_{\rm eff}$). The color represents overall metallicity ([M/H]). The solid lines indicate 5 Gyr PARSEC isochrones for reference.
\label{fig:kiel_diagram}}
\end{figure}

\begin{figure*}[htp!]
\centering
\includegraphics[width=170mm]{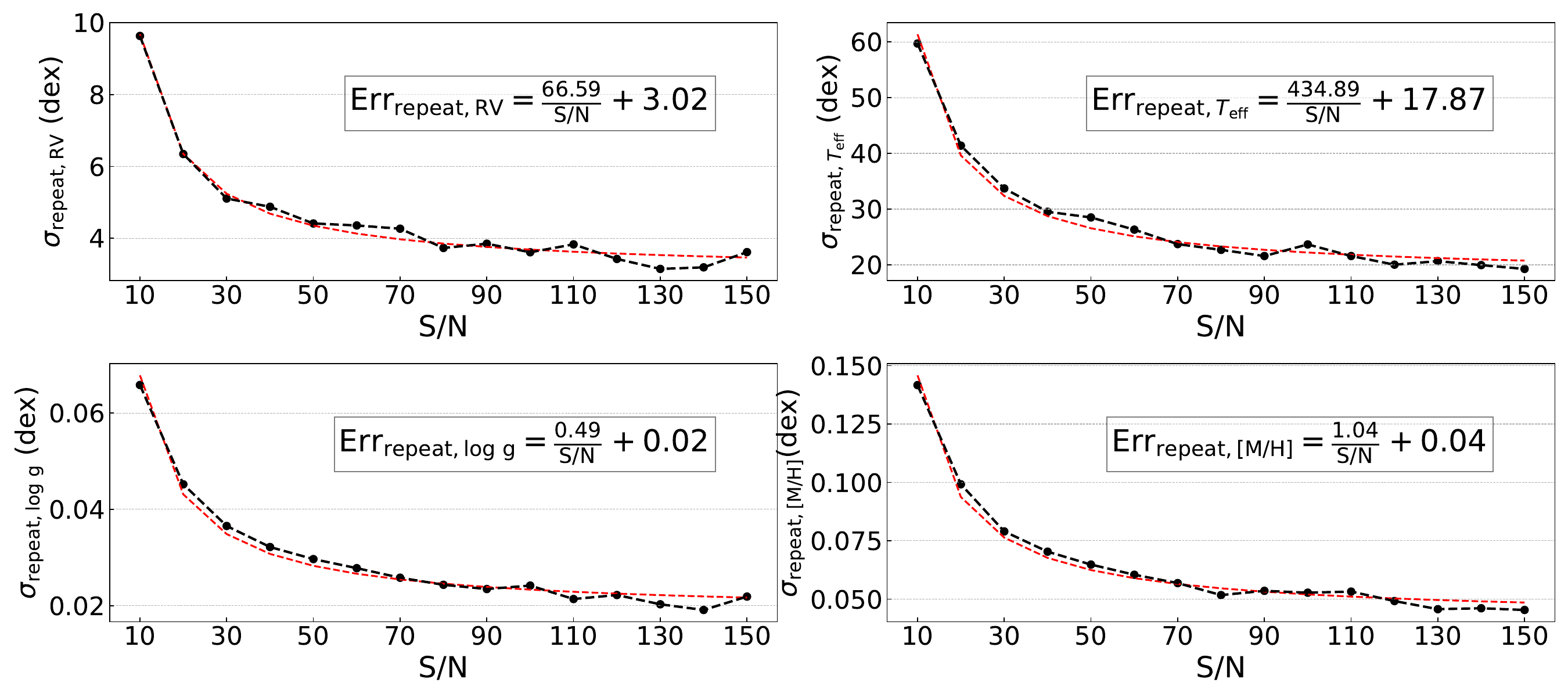}
\caption{Precision ($\rm Err_{\rm repeat}$) of RV, $T_{\rm eff}$, $\log g$, and [M/H] as a function of S/N, derived from repeated observations. Black dots show the standard deviation of repeated measurements in each S/N bin (width = 10). The red dashed curves show the best-fit inverse functions, with their expressions given in each panel. These relationships demonstrate how measurement precision improves with increasing S/N before stabilizing at S/N$>$100.
\label{fig:prec_func}}
\end{figure*}

As shown in Figure~\ref{fig:kiel_diagram}, the Kiel diagram illustrates the distribution of M dwarf stars in the parameter space. A comparison with the solid lines representing theoretical isochrones from the PARSEC evolutionary model indicates that the surface gravity and metallicity measurements generally align with the predictions of the isochrones.

The precision of parameter measurement is affected by both the spectral fitting errors and the quality of the observed spectra. Multiple LAMOST observations of the same sources under different conditions allow us to evaluate parameter measurement precision as a function of spectral S/N. As a result, the total error for a stellar parameter is calculated as the square root of the sum of the squares of the repeated measurement error (Err$_{\text{repeat}}$) and the fitting error (Err$_{\text{fitting}}$).

\begin{equation}\label{equa:err}
    \rm Err_{\text{total}} = \sqrt{\text{Err}_{\text{repeat}}^2 + \rm Err_{\text{fitting}}^2}.
\end{equation}

For stars with more than three observations, we calculated the standard deviation of repeated measurements to quantify parameter precision. We then modeled the precision ($\text{Err}_{\text{repeat}, x}$) of each parameter $x$ using an inverse function of S/N, with coefficients $a$ and $b$.

\begin{equation}\label{equa:err_rep} 
\text{Err}_{\text{repeat},x} = \frac{a}{\text{S/N}}+b.
\end{equation}

Figure~\ref{fig:prec_func} shows how measurement precision improves with increasing S/N for RV, $T_{\rm eff}$, $\log g$, and [M/H]. We model these relationships using inverse functions of S/N (Equation \ref{equa:err_rep}). At S/N = 10, the precision is 9.7 km/s for RV, 61 K for $T_{\rm eff}$, 0.07 dex for $\log g$, and 0.14 dex for [M/H]. These values improve to 3.7 km/s, 22 K, 0.02 dex, and 0.05 dex respectively at S/N = 100, after which the precision improvements begin to plateau.

\subsection{Hertzsprung-Russell Diagram}

\begin{figure*}[htp!]
\centering
\includegraphics[width=170mm]{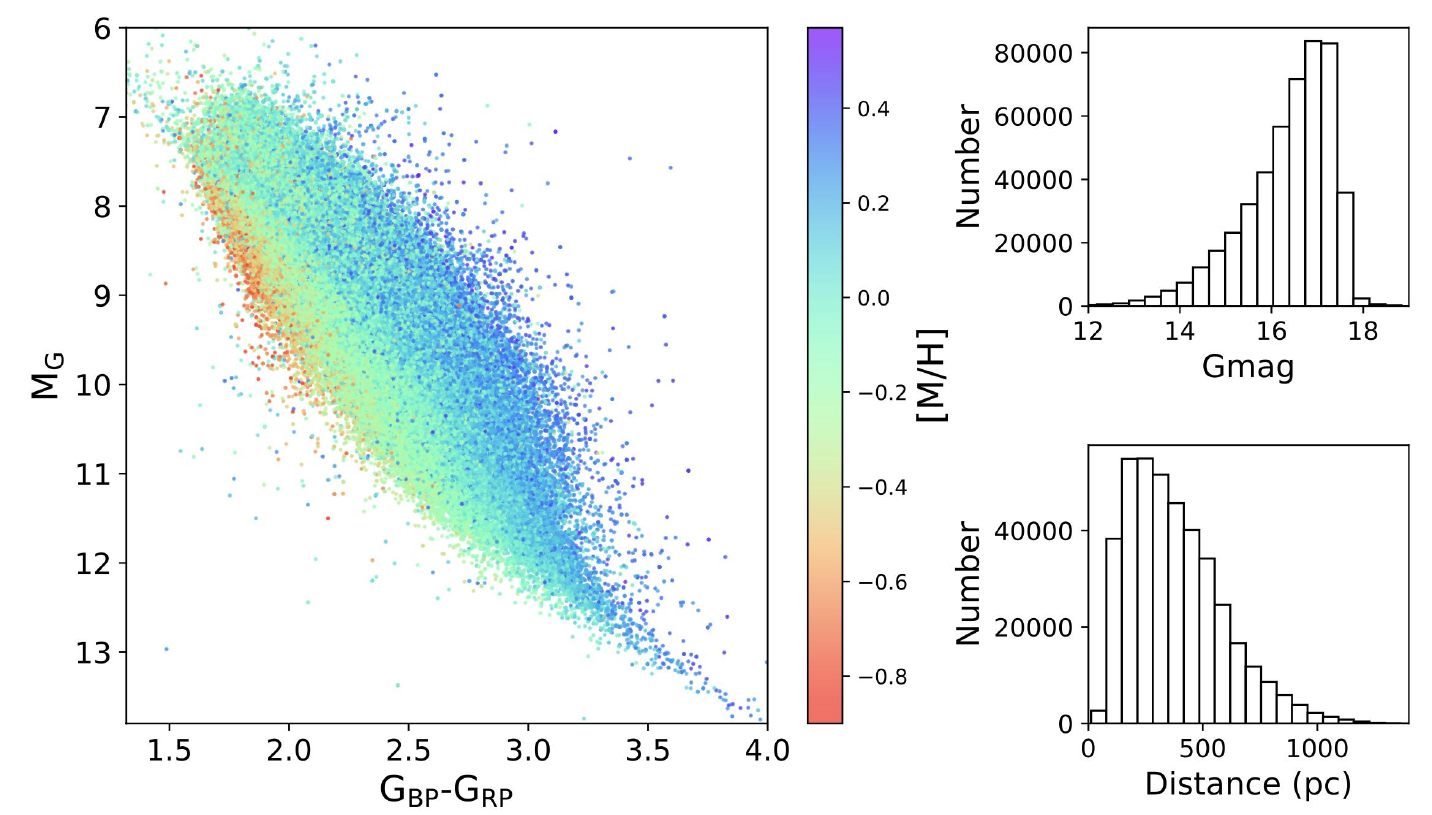}
\caption{Left panel: distribution of the M dwarf stars on the Gaia H-R diagram, color-coded by [M/H] measured in this study. Right panels: the upper and the lower display the distribution of Gaia G magnitude and distance of the M dwarf stars, respectively.
\label{fig:HR_diagram}}
\end{figure*}

With Gaia DR3 photometry available for the majority of our sample, we employ absolute $G$ magnitudes and BP$-$RP colors as luminosity and temperature proxies, respectively. The left panel of Figure \ref{fig:HR_diagram} presents the observational Hertzsprung-Russell (H-R) Diagram of our M dwarf sample, excluding variable stars and binary candidates (see ``VBFlag'' description in Section \ref{subsec:vbflag}). Given that most sources lie within 1 kpc (lower right sub-figure, Figure \ref{fig:HR_diagram}), we derived absolute $M_G$ magnitudes using Gaia DR3 parallaxes.

LAMOST's observational constraints restrict our sample to G-band magnitudes between 12 and 18 (lower right panel, Figure \ref{fig:HR_diagram}). The absolute magnitude distribution spans $M_G$ = 7-13, predominantly encompassing early to mid-type M dwarf stars.

\subsection{Activity Traced by H$\alpha$ Emission}

\begin{figure}[htp!]
\centering
\includegraphics[width=85mm]{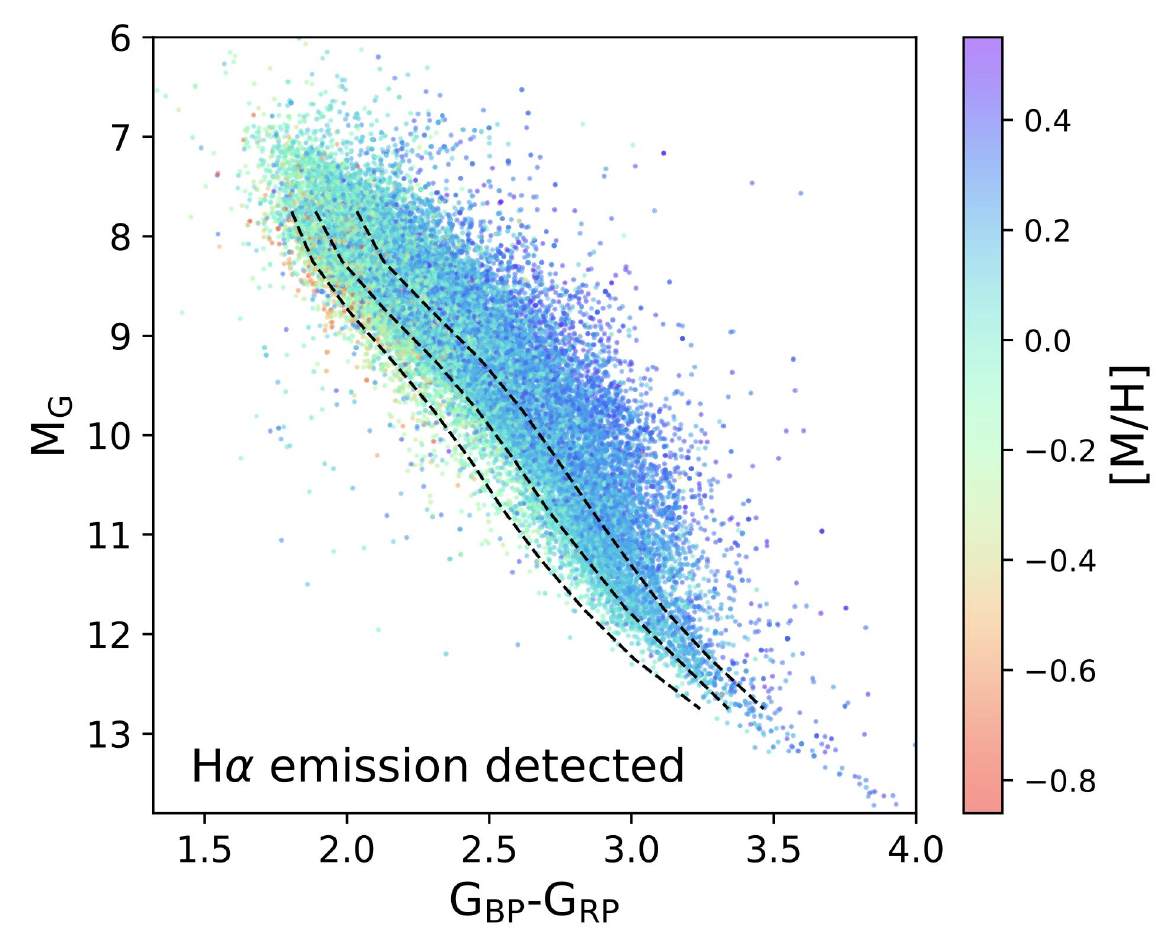}
\caption{Similar to the left panel of Figure \ref{fig:HR_diagram}, but displaying only the subset of the sample identified with H$\alpha$ emission. The three dashed lines represent 15, 50, 85 percentiles of the full sample shown in Figure \ref{fig:HR_diagram}, respectively.
\label{fig:Halpha}}
\end{figure}

H$\alpha$ emission serves as a fundamental indicator of chromospheric magnetic activity in M dwarf stars, first identified by \cite{1947ApJ...105...96J} who designated such stars as dMe. The fraction of H$\alpha$ active stars increases from early to late M dwarf stars, as demonstrated by both \cite{2008AJ....135..785W} and \cite{2019ApJS..243....2C} datasets, though this trend plateaus around $M_G\sim$15 (M7V). The interpretation of H$\alpha$ activity is complex, as \cite{1979ApJ...234..579C} and \cite{1986ApJS...61..531S} demonstrated that increasing chromospheric heating initially deepens H$\alpha$ absorption before transitioning to emission.

\cite{2017ApJ...834...85N} showed that M dwarf stars with masses around 0.30 $M_{\odot}$ typically exhibit H$\alpha$ activity when their rotation periods are less than 30 days, while this threshold extends to 80 days for 0.15 $M_{\odot}$ red dwarfs. \cite{2023ApJ...949...37P} further revealed that 74\% of active, fully convective M dwarf stars (0.1-0.3 $M_{\odot}$) are rapid rotators with periods under 2 days. Regarding variability, \cite{2020ApJ...905..107M} discovered that stronger H$\alpha$ emission correlates with more frequent flaring activity in fully convective M dwarf stars. \cite{2023AJ....166...63J} identified an H$\alpha$ deficiency zone at 10.3$<M_G<$10.8, linking reduced H$\alpha$ activity to slower rotation rates. Notably, \cite{2017ApJ...834...85N} suggested that for K and M dwarf stars, the chromospheric heating mechanism responsible for H$\alpha$ emission might operate independently of the underlying dynamo. Most fully convective M dwarf stars show H$\alpha$ variability that is not synchronized with rotation phase, suggesting that fixed photospheric features may not directly drive chromospheric H$\alpha$ variations. These findings collectively establish H$\alpha$ emission as a valuable tracer of magnetic activity in M dwarf stars, reflecting complex relationships among stellar rotation, internal structure, and evolutionary state.

To provide an H$\alpha$ emission indicator for our sample, we quantified stellar chromospheric activity by measuring the equivalent width (EW) of the H$\alpha$ emission line. The EW was calculated using the following equation:

\begin{equation}
    \text{EW} = -\int^{\lambda_2}_{\lambda_1}(1-\frac{F_{I\lambda}}{F_{C\lambda}})d\lambda,
\end{equation}
\noindent
where $F_{C\lambda}$ represents the local continuum flux in the line region, and $F_{I\lambda}$ denotes the spectral flux across the line profile. Following the standard procedure established by \cite{2004AJ....128..426W,2011AJ....141...97W}, we measured the EW within a 14 \AA \ \ window centered on the H$\alpha$ line ($\pm$7 \AA \ \ \ from the line center). A star is considered chromospherically active if its H$\alpha$ EW exceeds 0.75 \AA. Then we defined an ``Activity'' flag in our catalog, assigning a value of 1 to stars with EW$>$0.75 \AA \ \ and 0 to those below this threshold. 

In Figure \ref{fig:Halpha}, the distribution of H$\alpha$ active M dwarf stars (33,623 objects) on the H-R diagram reveals that active stars are notably elevated above the main sequence, which is consistent with the literature (e.g., \citealt{2023AJ....166...63J}). These stars show unusually high metallicities, which could be attributed to several factors. First, this might reflect the age-metallicity correlation, as younger stars tend to be both more metal-rich and more active. Second, some of these stars could be tight binary systems, where the companion affects both the activity levels and spectral features. Third, a portion of these elevated, high-activity sources might be pre-main sequence stars, whose different spectral characteristics and CMD positions could lead to apparently higher metallicity estimates. Additionally, residual activity effects might persist even after H$\alpha$ masking. Given these complexities, we suggest that their derived parameters should be used with caution.

\subsection{Subclass and Spectral Subtype}\label{subsec:spt}
Analysis of the Hertzsprung-Russell diagram reveals a number of bright stars with $M_G<$8, which exceed typical M dwarf luminosities~\citep{2024ARA&A..62..593H}. These are likely misclassified late-type K dwarf stars, as spectral classification systems occasionally vary by 1$\sim$2 subtypes. As described in Section \ref{sec:data}, our initial M dwarf candidates were classified using LAMOST full-spectrum template fitting. Therefore, we implemented an alternative classification method for comparison. To differentiate between the two, the original LAMOST classification is referred to as the ``spectral subclass'', while the new classifications are termed ``spectral subtype (SpT)''.

LAMOST observational constraints limit our sample primarily to stars earlier than M6, as late-type M dwarf stars' spectral energy distributions peak in the near-infrared (M dwarf stars with $T_{\rm eff}\sim$2,900 K peak at 1 $\mu$m). We adopt the SpT--TiO5 relation, which was originally defined by \cite{1995AJ....110.1838R} and applied to SDSS M dwarf stars by \cite{2007AJ....133..531B}:
\begin{equation}
\text{SpT} = -10.775\times\text{TiO5} + 8.2, \ \ \ \ \ \sigma=\pm0.5,
\end{equation}
\noindent
where TiO5 index is defined as

\begin{equation}
\text{TiO5}=\frac{F_W}{F_{\rm cont}},
\end{equation}
\noindent
with $F_W$ representing the mean flux of TiO5 band head (7126-7135 \AA) and $F_{\rm cont}$ the mean flux of its pseudo-continuum (7042-7046 \AA). This relation applies to spectral subtypes from $-$2 to 7, where $-$2 denotes K5, $-$1 denotes K7, and 0 through 7 correspond to M0 through M7.

\begin{figure}[htbp!]
\centering
\includegraphics[width=90mm]{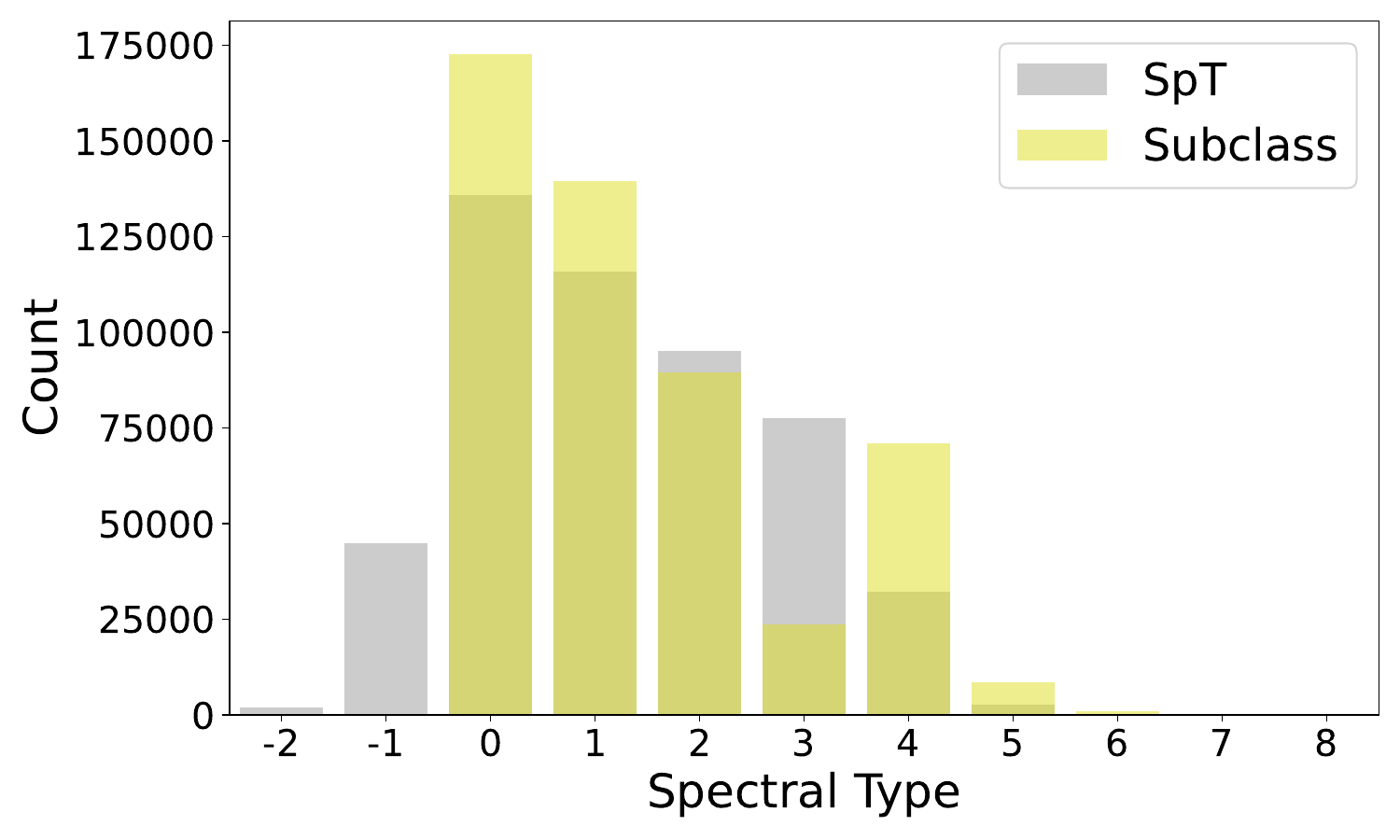}
\caption{This histogram compares the spectral subclass distribution from the LAMOST pipeline with the spectral subtype (SpT) distribution derived using the empirical TiO5--SpT relationship, as described in Sec.\ref{subsec:spt}.
\label{fig:spt}}
\end{figure}

As a result, a systematic offset of 0.18 subclass with a standard deviation of 0.74 subclass was identified between the two classifications across the sample. Figure \ref{fig:spt} illustrates this comparison, suggesting potential contamination by misclassified late-K type dwarf stars. We included both classifications in our catalog, with $-$9999 indicating cases where SpT calculation was impossible due to spectral quality limitations.

\subsection{Candidate Variable Stars and Binaries}\label{subsec:vbflag}

Given that our theoretical models are based on single-star assumptions, parameter predictions for unresolved binary systems and variable stars would have reduced accuracy. Therefore, we initially applied a RUWE$<$1.4 criterion to our sample selection, and additionally, we provided a flag for variable star candidates and binary candidates in our catalog through cross-matching with various catalogs.

For variable stars, we primarily used the Gaia DR3 Vari Summary Catalog, which contains 11,754,237 objects with VarFlag=``VARIABLE''. This catalog incorporates data from the Gaia vari\_classifier\_result table (\citealt{2023A&A...674A..14R}; 9,976,881 sources) and Gaia DR3 × literatures (\citealt{2023A&A...674A..22G}; 4.9 million objects), yielding 24,276 matches from our sample.

For binary stars, we cross-matched with the following catalogs:

(1) Gaia EDR3 200 pc binary candidates catalog~\citep{2023AJ....166..218M}: 2,585 matches from 235,269 pairs within 2.5'';

(2) Gaia DR3 eclipsing binary catalog~\citep{2023A&A...674A..16M}: 447 matches from 2,184,477 objects, including entries from the Gaia non-single stars catalog (810,000 objects) and DR3 DPAC orbital solutions (86,918 objects);

(3) APOGEE SB2 catalog~\citep{2021AJ....162..184K}: 240 matches from 7,273 objects;

(4) 2021 final version of SB9 (the 9th Catalogue of Spectroscopic Binary orbits; \citealt{2004A&A...424..727P}): 1 match from 4,021 objects;

(5) DBECat (Detached Eclipsing Binary Stars;~\citealt{2015ASPC..496..164S}): 4 matches;

(6) LAMOST MRS SB2 catalogs: 336 matches from \cite{2023ApJS..266...18Z} which contains 36,470 candidates and 102 matches from \cite{2024MNRAS.527..521K} which contains 12,426 candidates.

As a result, we implemented a ``VBFlag'' in our catalog: 1 for variable candidates (24,276 objects), 2 for binary candidates (3,455 objects), and 0 for all others, indicating no detected variability or binarity based on our cross-matching criteria.

\subsection{Description of the Catalog}
Table \ref{tab:desrcription} presents a comprehensive data description of our value-added M dwarf catalog, containing 23 columns of stellar parameters and observational measurements. The catalog combines data from LAMOST spectroscopic observations and Gaia DR3 astrometric measurements, along with our derived parameters. Each star is identified by a unique ID and each LAMOST spectrum is indicated by a unique identifier (obsid). Positional information includes J2000 coordinates (RA and Dec). Spectral quality is quantified by signal-to-noise ratios in both r and i bands (SNRr, SNRi). Two spectral classification systems are provided: the original LAMOST pipeline subclass and our derived spectral subtype (SpT) based on the TiO5 index.

Photometric data from Gaia DR3 includes G-band mean magnitude (Gmag) and integrated BP and RP magnitudes (BPmag, RPmag). Astrometric measurements comprise parallax (Plx) and its associated error (e\_Plx), along with the Gaia Renormalized Unit Weight Error (RUWE). Our derived parameters include radial velocity (RV) with typical uncertainties (e\_RV), effective temperature (Teff), surface gravity (Logg), and overall metallicity ([M/H]), each accompanied by their respective uncertainties. Two additional flags indicate H$\alpha$ emission activity (Activity) and potential variability or binarity (VBFlag).

\section{Comparison}\label{sec:comp}

\begin{figure*}[htp!]
\includegraphics[width=180mm]{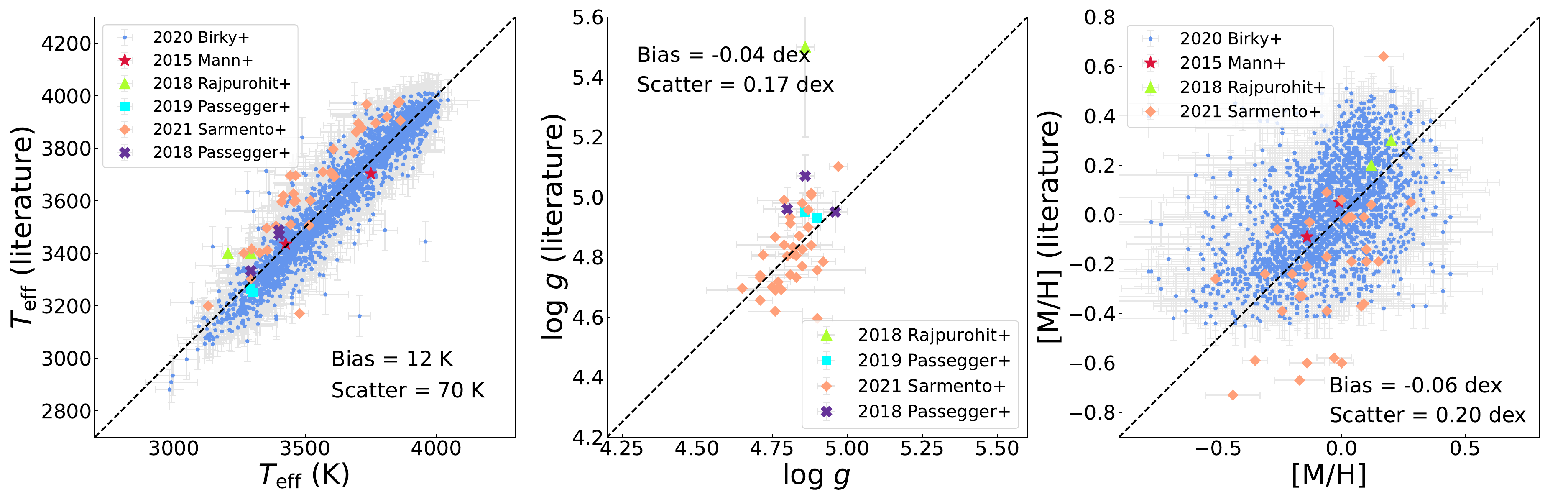}
\caption{Comparison of three basic parameters between this work and the available results from the literature, including \cite{2015ApJ...804...64M}, \cite{2018A&A...615A...6P}, \cite{2018A&A...620A.180R}, \cite{2019A&A...627A.161P}, \cite{2020ApJ...892...31B}, and \cite{2021A&A...649A.147S}. The total sample size available for comparison, along with the bias and scatter for each parameter, is indicated in the upper left corner of each panel. The dashed lines represent the 1:1 reference lines.
\label{fig:highres}}
\end{figure*}

\begin{figure*}[htb!]
\includegraphics[width=180mm]{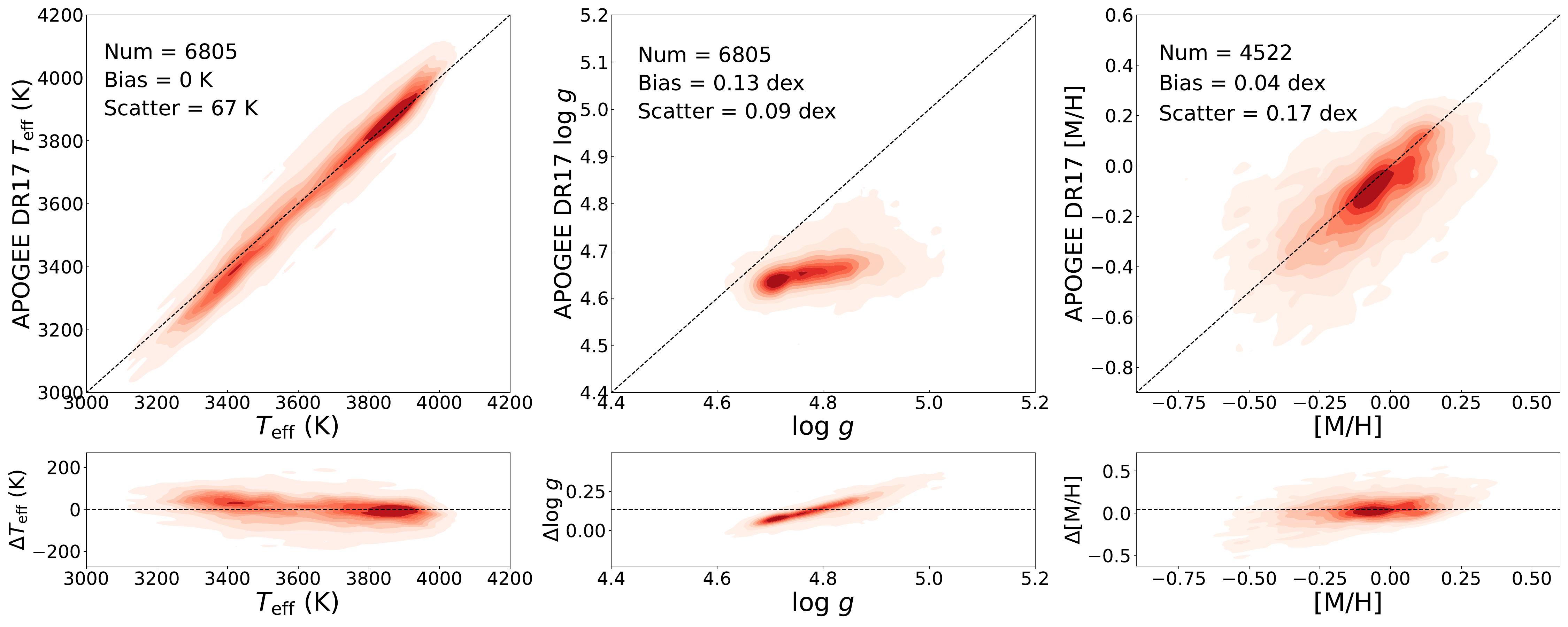}
\caption{Comparison of three basic parameters derived in this work with results from APOGEE DR17. The diagrams are color-coded by number density. For this comparison, DR17 spectroscopic temperatures and calibrated $\log g$ values are employed. The total sample size available for comparison, as well as the bias and scatter of the differences for each parameter, are shown in the upper left corner of each panel. The dashed lines represent the 1:1 reference lines. As APOGEE DR17 only provides [M/H] values for samples with $T_{\rm eff}>$ 3500 K, the sample size used in the right panel for metallicity comparison is smaller than those used in the other two panels.\label{fig:apogee}}
\end{figure*}

In this section, we evaluated the accuracy of the parameters by comparing our results with established datasets, including spectral parameters from the literature, the APOGEE survey, and other studies on LAMOST M dwarf parameterization. To further validate the results, metallicity measurements were cross-checked using FGK+M binaries.

\subsection{Comparison with Results from Literature}

\begin{figure*}[htp!]
\centering
\includegraphics[width=180mm]{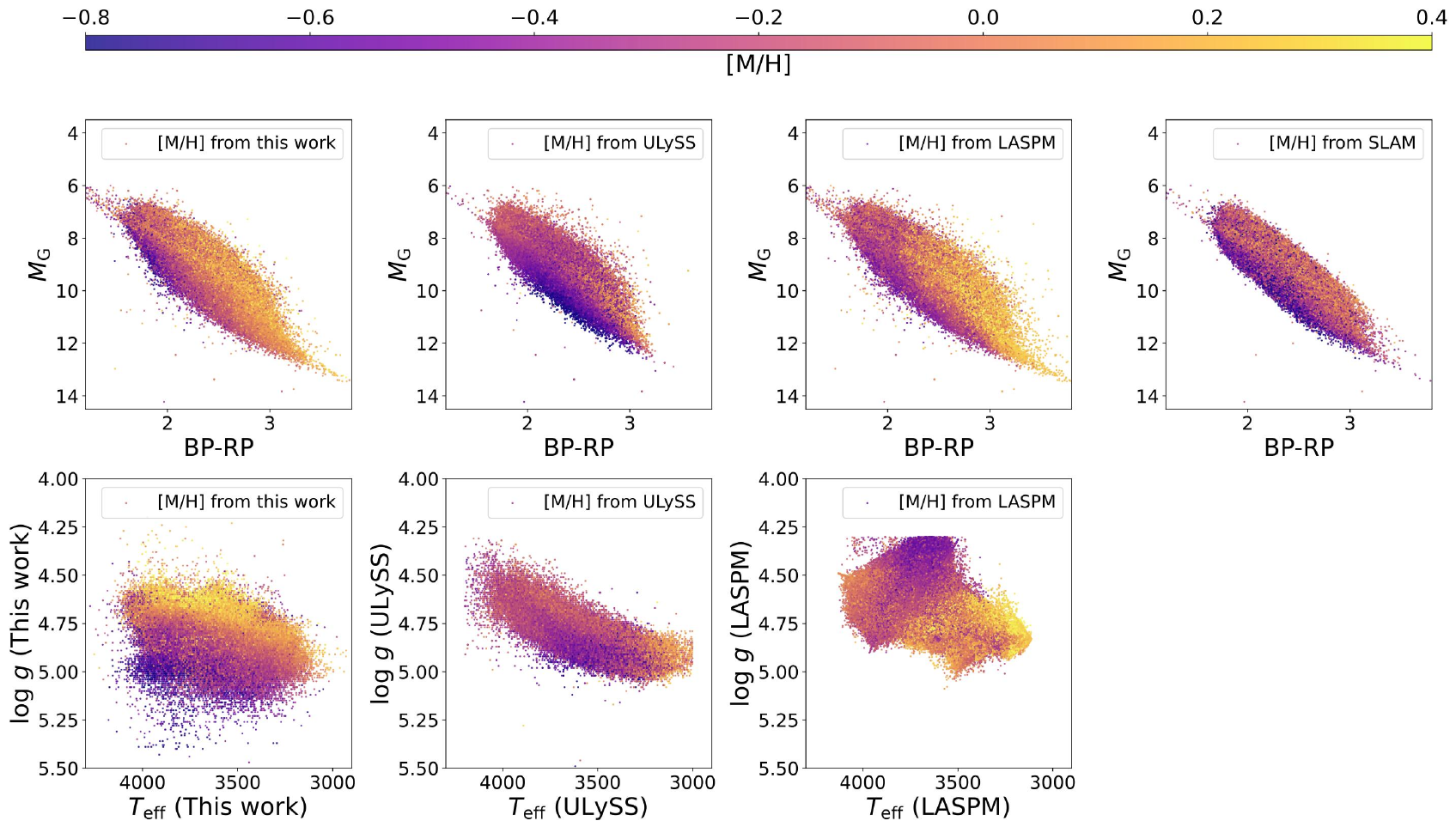}
\caption{H-R Diagrams and Kiel diagrams of LAMOST M dwarf parameters derived from 4 different works: this study, ULySS~\citep{2022ApJS..260...45D}, LASPM~\citep{2021RAA....21..202D}, and SLAM~\citep{2021ApJS..253...45L}. The parameters from LASPM are the latest DR11 version~\citep{2024arXiv241016730D}. Note that in the figures, we uniformly labeled both the [Fe/H] derived from empirical templates and the overall metallicity [M/H] obtained from theoretical models as [M/H]. Each diagram is color-coded by the metallicity derived in each work, using a consistent colormap across all diagrams.
\label{fig:lamost_works}}
\end{figure*}

\begin{figure*}[htbp!]
    \centering
    \includegraphics[width=0.33\textwidth]{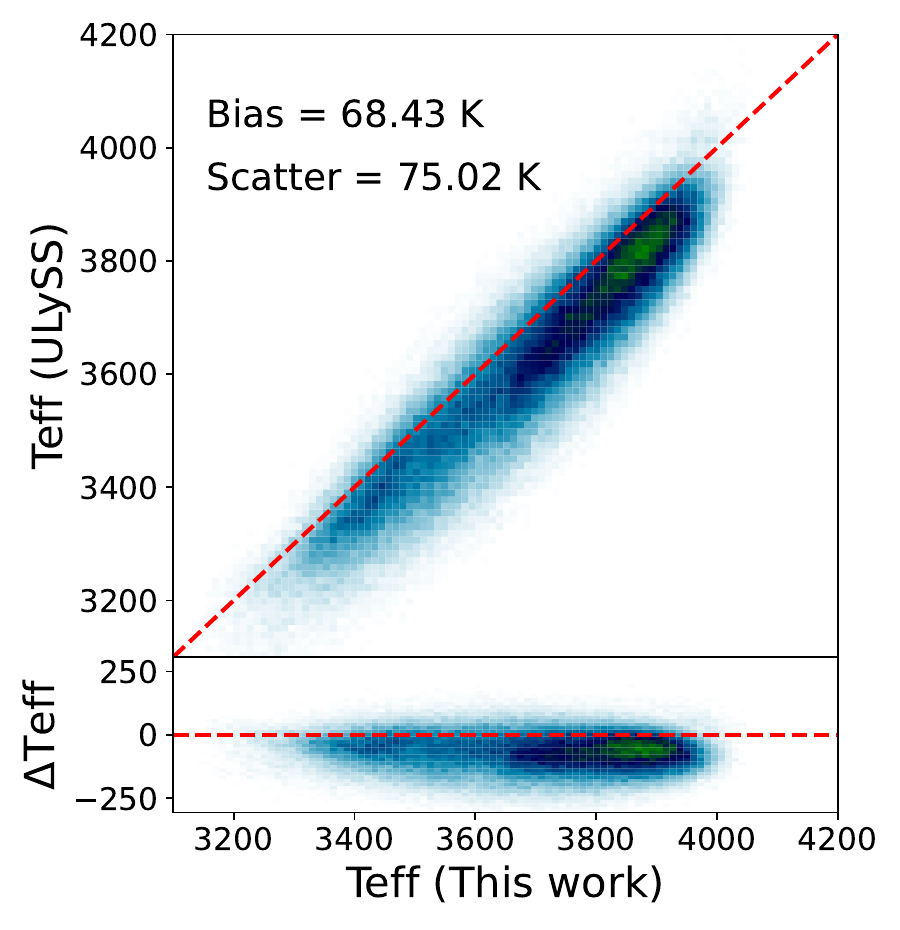}
    \includegraphics[width=0.33\textwidth]{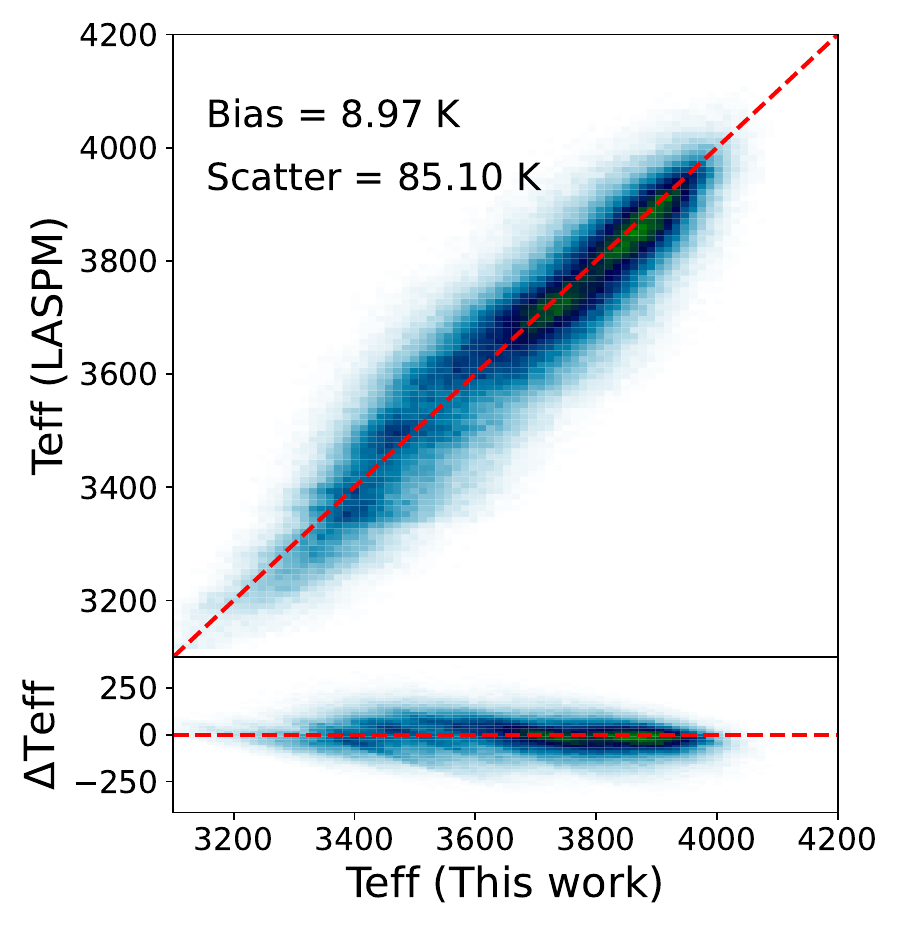}
    \includegraphics[width=0.33\textwidth]{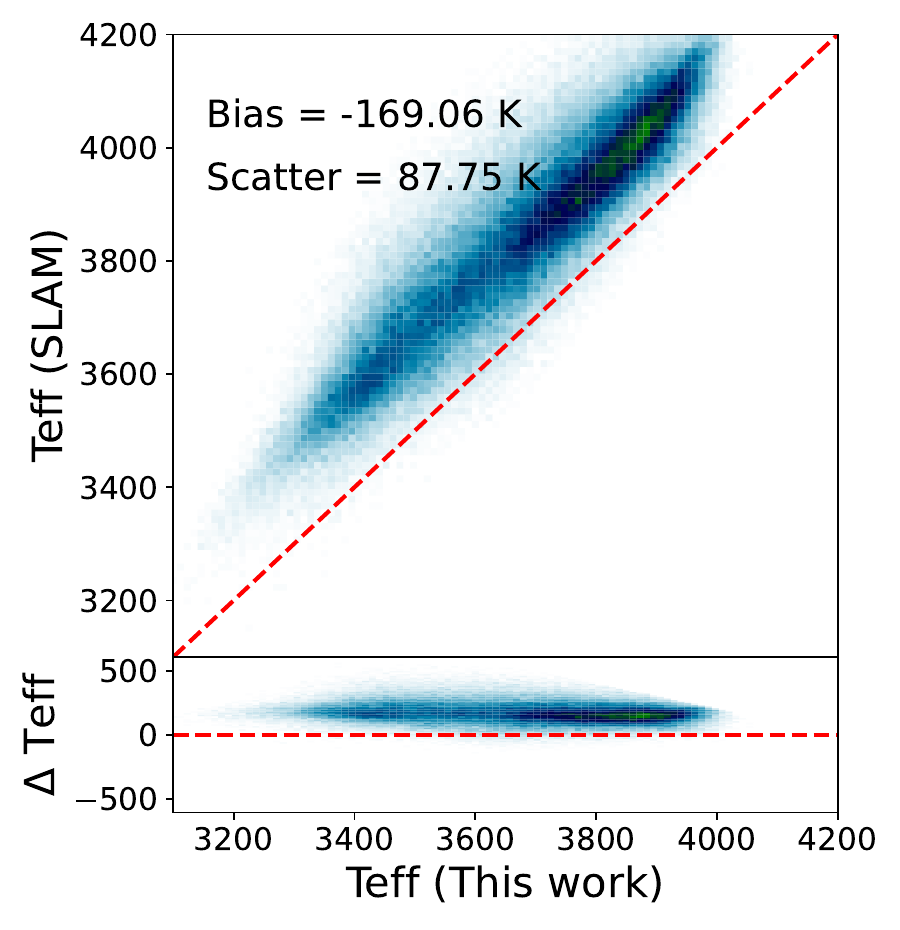}
    \includegraphics[width=0.33\textwidth]{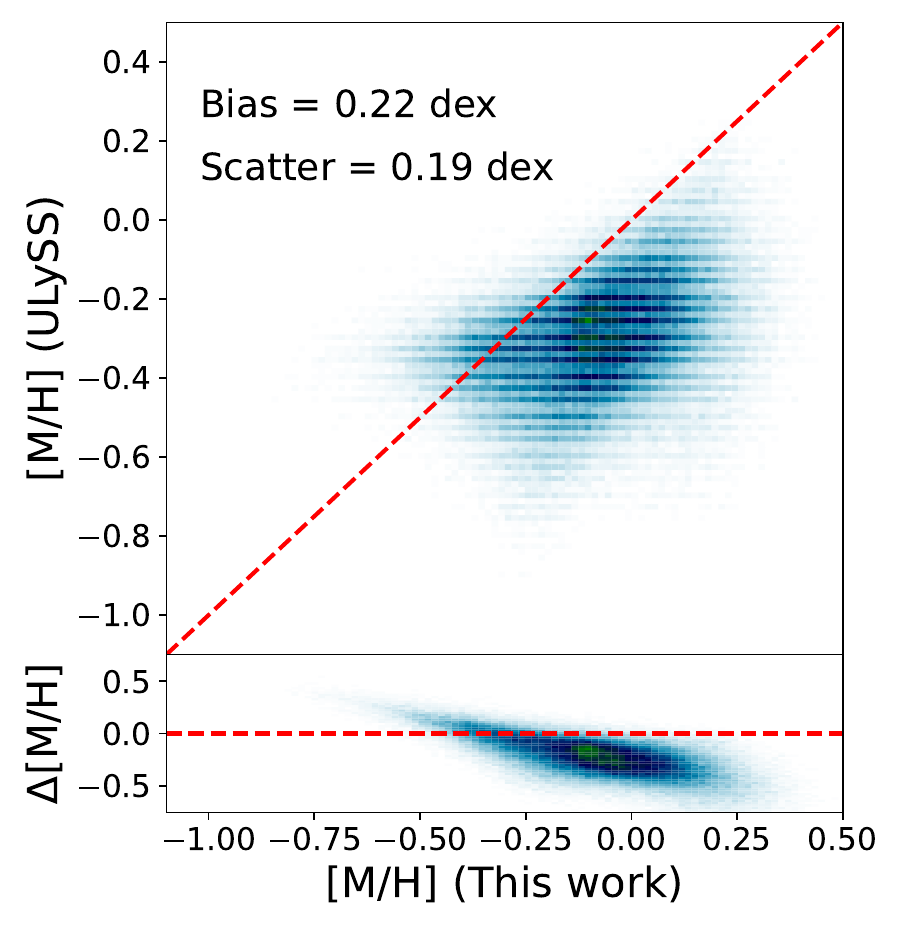}
    \includegraphics[width=0.33\textwidth]{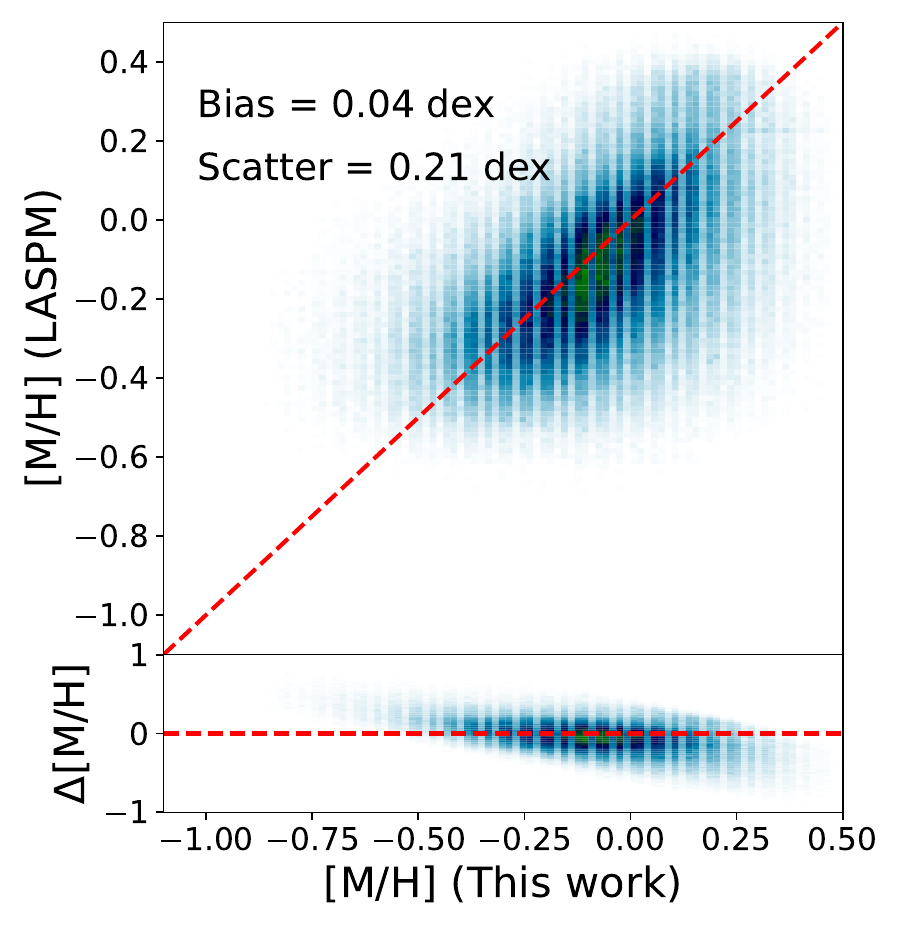}
    \includegraphics[width=0.33\textwidth]{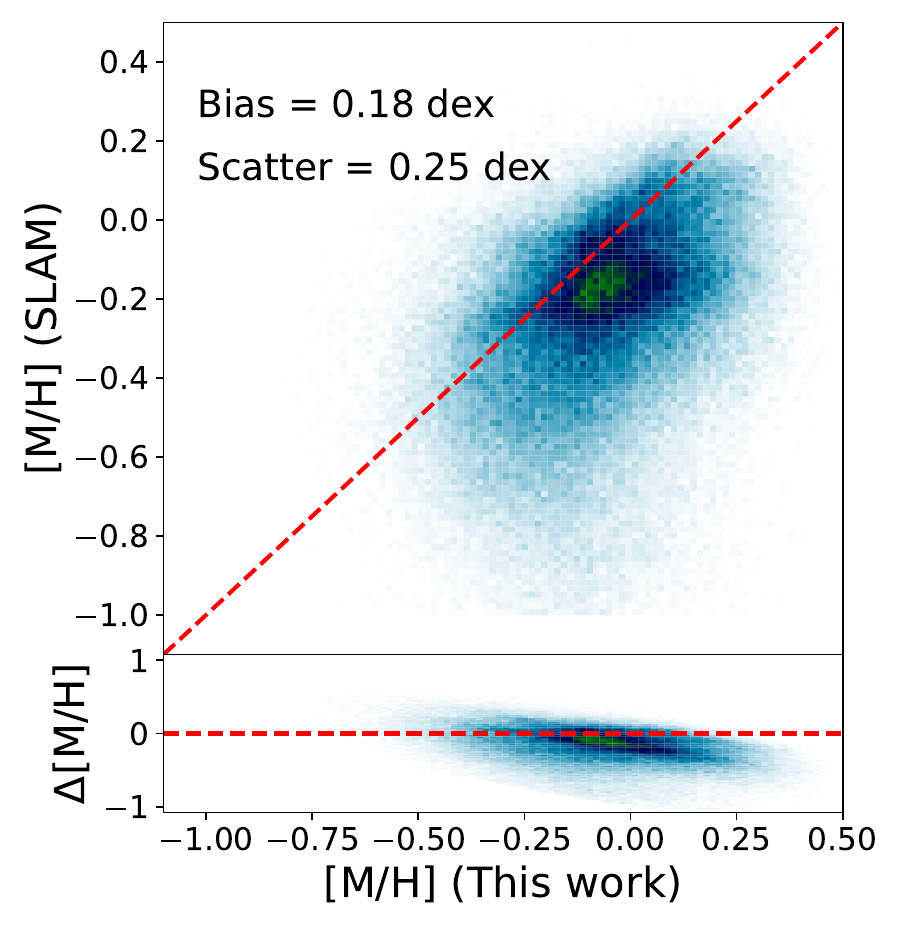}
    \includegraphics[width=0.33\textwidth]{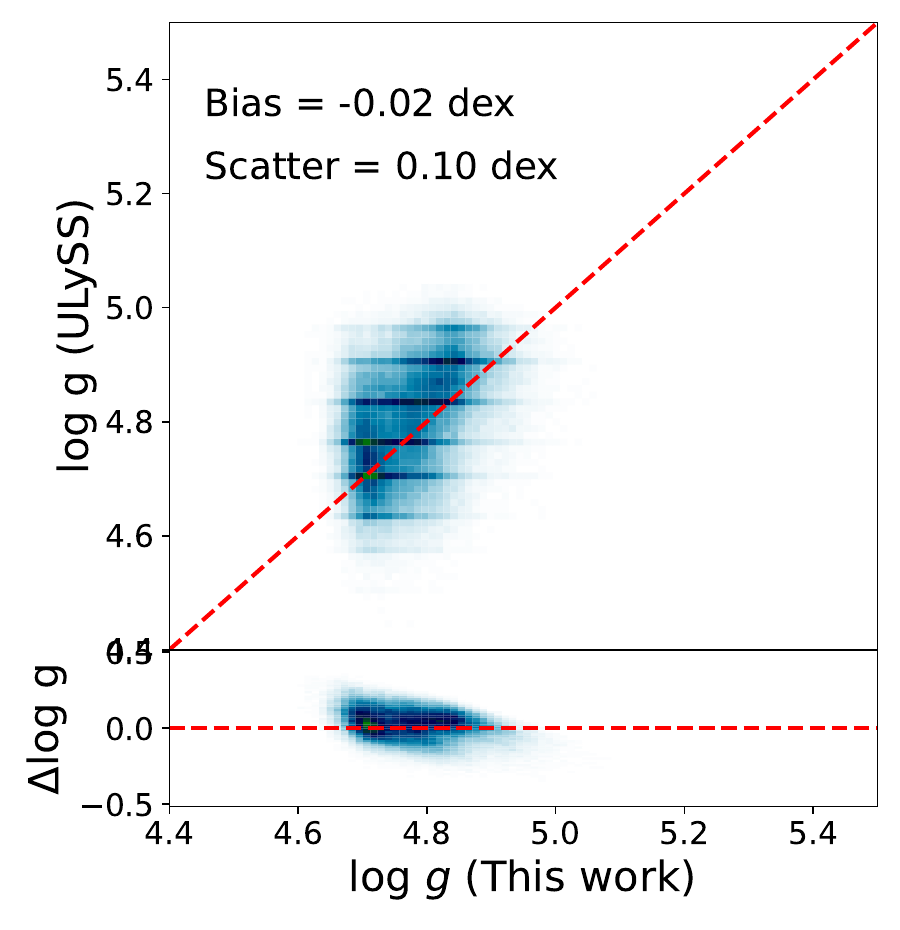}
    \includegraphics[width=0.33\textwidth]{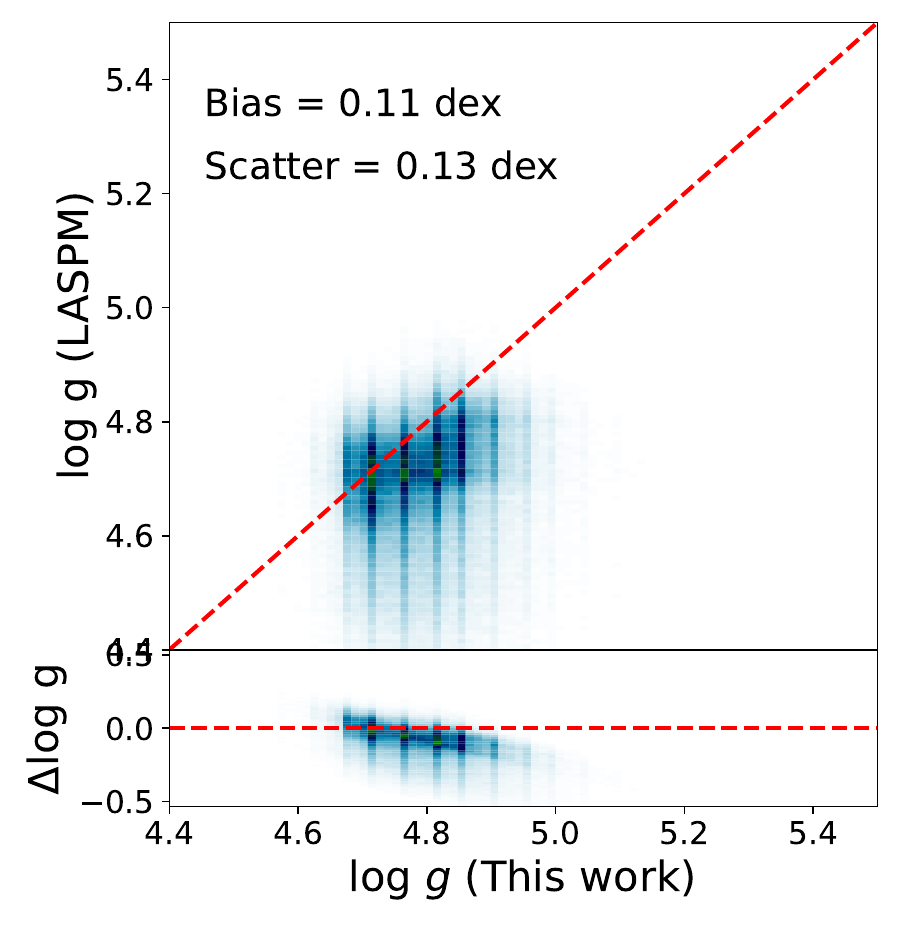}
    \includegraphics[width=0.33\textwidth]{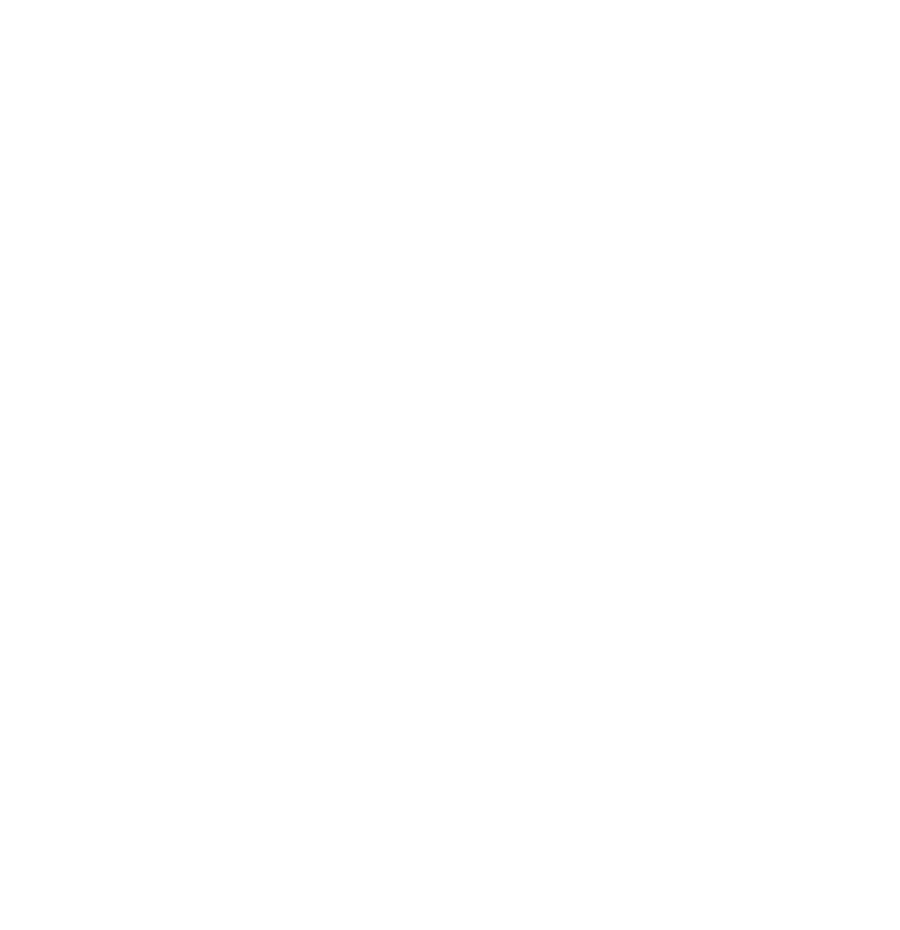}
    \caption{Comparison of three basic parameters derived in this work with results from ULySS~\citep{2022ApJS..260...45D}, LASPM~\citep{2021RAA....21..202D,2024arXiv241016730D}, and SLAM~\citep{2021ApJS..253...45L} respectively. The bias and scatter for each parameter are indicated in each panel. The diagrams are color-coded by number density.
    \label{fig:comp_lamost_works}}
\end{figure*}

Figure \ref{fig:highres} presents a comprehensive comparison of our derived fundamental parameters ($T_{\rm eff}$, $\log g$, and [M/H]) with values from several key literature sources spanning 2015-2021. The effective temperature comparison in the left panel shows generally good agreement across the M dwarf temperature range (3000-4000 K), with a small systematic offset of 12 K and a scatter of 70 K. When restricting to higher quality spectra (S/N$>$50), the scatter reduces to 47 K with a similar offset of 16 K. This level of consistency is noteworthy given the diverse methodologies employed by different studies. Surface gravity measurements exhibit a tight correlation with literature values, showing a bias of $-$0.04 dex and a scatter of 0.17 dex, which improves to -0.08 dex and 0.12 dex respectively for high S/N spectra, though the comparison sample is smaller and primarily concentrated around $\log g \sim 5.0$. The metallicity comparison reveals broader scatter (0.20 dex) and a slight systematic offset of $-$0.06 dex, with the largest discrepancies appearing at the metallicity extremes. For high S/N spectra, the scatter improves to 0.14 dex, while maintaining a similar metallicity range in the comparison sample.

The \cite{2020ApJ...892...31B} sample provides the most extensive overlap with our measurements. Their study employed The Cannon on a dataset of 5875 APOGEE M dwarf stars, using a training set with optically derived and bolometric-calibrated temperatures. Specialized studies like \cite{2021A&A...649A.147S} offer valuable benchmark comparisons for surface gravity determinations through their comparison of APOGEE observations with a new set of synthetic spectra. These results demonstrate that our parameter determinations achieve comparable precision to previous studies while substantially expanding the sample size of characterized M dwarf stars.

\subsection{Comparison with APOGEE DR17}
Figure \ref{fig:apogee} presents a detailed comparison between our derived parameters and those from APOGEE DR17, with density plots showing the distribution of parameter differences. 

For effective temperature, the comparison of 6,805 common objects shows excellent agreement, with no systematic bias and a scatter of only 67 K across the entire temperature range (3000-4200 K). The $\log g$ comparison reveals a systematic offset of 0.13 dex and a scatter of 0.09 dex, with our measurements consistently yielding higher values than APOGEE DR17 (calibrated) surface gravities. This offset is pronounced around $\log g \sim 4.7$, though the tight scatter suggests a systematic rather than random difference between the two surveys' calibration scales. The metallicity comparison, limited to 4,522 stars with $T_{\rm eff}$$>$3500 K in APOGEE DR17, shows good agreement with a small bias of 0.04 dex and a scatter of 0.17 dex. The bottom panels demonstrate that these parameter differences remain stable across their respective ranges, with no significant trends that might indicate systematic biases in specific parameter regimes. This comparison with APOGEE's high-resolution spectroscopic measurements validates the reliability of our parameter determinations.

\subsection{Comparison with Other LAMOST M Dwarf Parameterization Works}

In Figures \ref{fig:lamost_works} and \ref{fig:comp_lamost_works}, we compared our parameter determinations with three other LAMOST M dwarf studies that employed different methodological approaches: \cite{2021ApJS..253...45L} employed SLAM (Stellar LAbel Machine), a data-driven model based on support vector regression, trained on both APOGEE DR16 labels (for $T_{\rm eff}$ and [M/H]) and BT-Settl synthetic spectra (for $T_{\rm eff}$ only) to estimate stellar parameters of M dwarf stars in LAMOST DR6; \cite{2022ApJS..260...45D} used the ULySS package to perform $\chi^2$ minimization between LAMOST DR8 M-type observed spectra and model spectra generated from the MILES \citep{2016A&A...585A..64S} interpolator; LASPM \citep{2021RAA....21..202D,2024arXiv241016730D} estimated stellar atmospheric parameters using $\chi^2$ minimization to find the five best-matching templates and combines them linearly, initially using PHOENIX BT-Settl theoretical spectra as references (2021, applied to LAMOST DR6 and DR7), and later evolving to use empirical LAMOST spectral templates derived through label transfer from Gaia EDR3 (2024, applied to DR10 and DR11).

As these studies utilized different datasets for parameter estimation, we performed cross-matching to obtain common sources for comparison. Both ULySS and LASPM provided predictions for three parameters ($T_{\rm eff}$, $\log g$, [M/H]), with LASPM results from DR11 being used in our comparison. For SLAM, we compared with their $T_{\rm eff}$ and [M/H] results derived from APOGEE DR16 labels. The H-R diagrams in the upper panels of Figure \ref{fig:lamost_works} show that our Cycle-StarNet-based approach yields a broader metallicity distribution than ULySS and SLAM, particularly for cooler stars ($T_{\rm eff}<3500$ K), while showing similar patterns to LASPM's results. The Kiel diagrams in the lower panels further illustrate these methodological differences through the varied distributions in the three-dimensional parameter space.

The quantitative comparisons in Figure \ref{fig:comp_lamost_works} highlight systematic differences arising from these methodological choices. Our effective temperatures are systematically cooler than ULySS (bias = 68.43 K) but warmer than SLAM (bias = $-$169.06 K), reflecting the different temperature scales of theoretical models versus empirical calibrations. The closest temperature agreement is with LASPM (bias=8.97 K). Surface gravity shows distinct patterns: a tight correlation with ULySS (scatter=0.10 dex) but more dispersed relationships with LASPM (scatter=0.13 dex), highlighting the challenges in gravity determination from spectra. The metallicity scales show notable systematic offsets (ULySS: 0.22 dex, SLAM: 0.18 dex) except with LASPM (0.04 dex). These systematic differences highlight the complexity in determining atmospheric parameters of M-type stars, as different methodological choices -- theoretical models, empirical libraries, or machine learning -- lead to varying parameter scales while maintaining similar relative stellar parameter rankings.

\subsection{Comparison of Metallicity in FGK+M Binaries}\label{subsec:binary}

\begin{figure}[htp!]
\includegraphics[width=82mm]{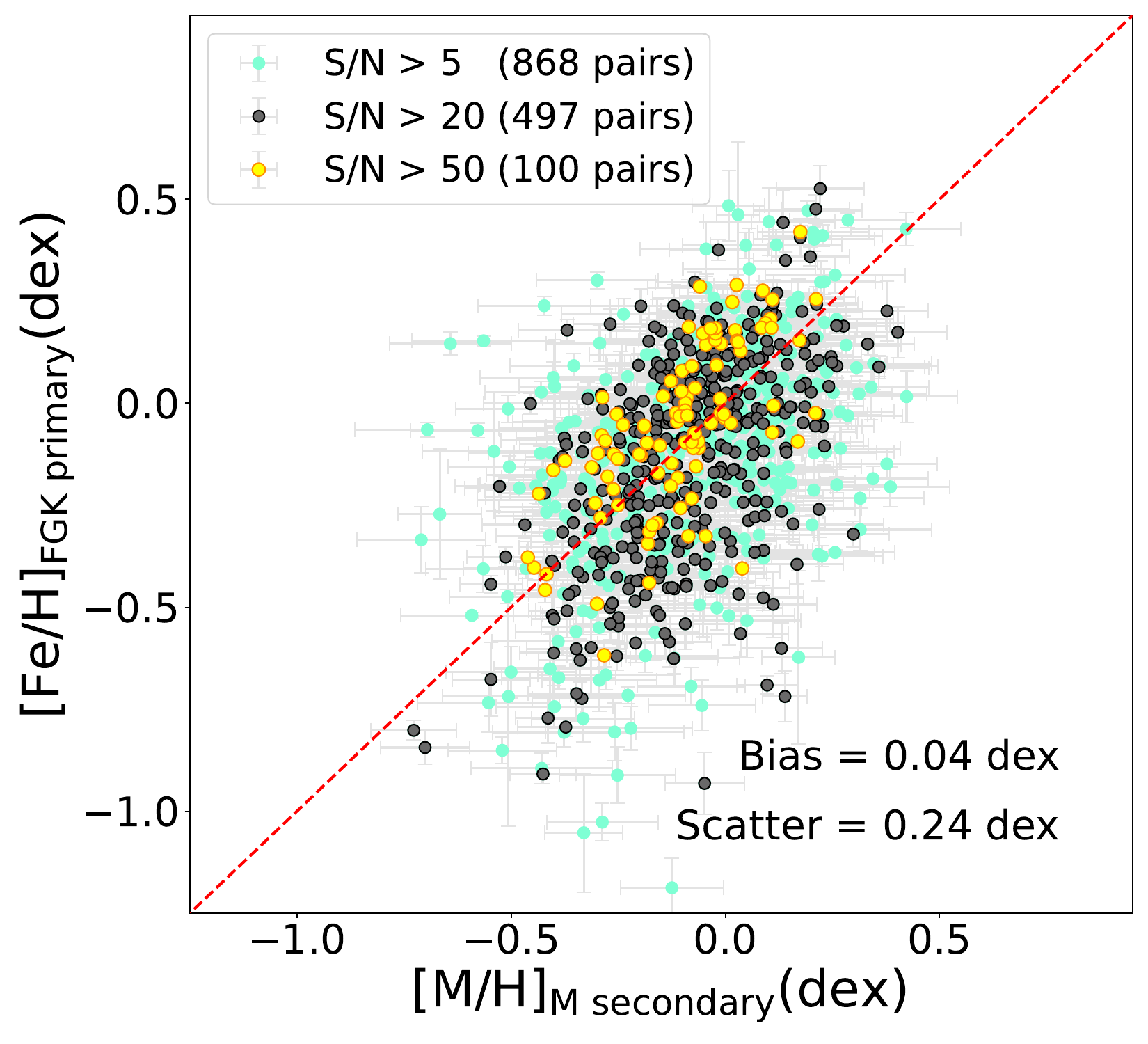}
\caption{Metallicity comparison between FGK primaries and M dwarf secondaries in wide binary systems identified from \cite{2021MNRAS.506.2269E}. The FGK primary metallicities are from LASP, while M dwarf values are from this study. Colors indicate different S/N thresholds, with sample sizes decreasing from 868 pairs (S/N$>$5) to 100 pairs (S/N$>$50). The bias and scatter for all 868 pairs are shown. The red dashed line represents the 1:1 reference line.
\label{fig:fgkm}}
\end{figure}

Wide binary systems, where component stars form from the same molecular cloud, provide excellent testbeds for metallicity calibration. The well-determined metallicities of FGK primaries serve as benchmarks for validating M dwarf metallicity measurements.

We analyzed M dwarf metallicities using their FGK binary companions as calibrators. Our sample derives from the \cite{2021MNRAS.506.2269E} catalog, which contains 1,871,594 wide binary candidates within 1 kpc from Gaia EDR3. We cross-match this catalog with LAMOST data and apply three selection criteria:
\begin{enumerate}
\item R\_chance\_align $<$ 0.01 (99\% binding probability)
\item pairdistance $>$ 3 arcseconds (avoiding fiber contamination)
\item $|\Delta \text{RV}|<$ 10 km s$^{-1}$
\end{enumerate}

These criteria yield 868 binary pairs. We compared LASP measurements for FGK primaries with our M dwarf measurements under different S/N threshold limitations, as Figure \ref{fig:fgkm} shows. Note that LASP provides [Fe/H] rather than [M/H], based on the library of standard observed spectra. For the full sample (S/N$>$5), we find a small systematic offset (bias = 0.04 dex) and a scatter of 0.24 dex between FGK primaries and M dwarf secondaries, with the metallicity distribution spanning [-1.2, 0.5] dex (5th to 95th percentiles). At S/N$>$20, the bias reduces to 0.02 dex with a scatter of 0.21 dex, while maintaining a similar metallicity range of [-0.9, 0.5] dex. For the highest quality spectra (S/N$>$50), we achieve a scatter of 0.15 dex with a bias of $-$0.04 dex. While this subsample of 100 pairs shows a somewhat narrower metallicity range of [-0.7, 0.2] dex, the systematic reduction in scatter with increasing S/N, together with the maintained coverage of metallicity space at intermediate S/N, suggests that the improved consistency primarily reflects enhanced measurement precision rather than sample selection effects.

\section{Summary}\label{sec:con}

Traditional spectroscopic analysis of M dwarf stars faces significant challenges due to imperfect theoretical models, parameter degeneracy, and difficulties in normalizing molecular band-dominated spectra. These challenges are particularly acute for low-resolution surveys like LAMOST, where 60\% of spectra have S/N $<$ 20 in the r-band. To address these limitations, we implemented Cycle-StarNet, an unsupervised domain adaptation method that learns to transform between synthetic and observed spectral domains while preserving physical stellar properties.

Our methodology incorporated stellar evolution constraints and parameter ranges from high-resolution studies to ensure physically meaningful results. The training process reduced the synthetic gap, as evidenced by residual analysis showing improvement from over 2 times the flux uncertainty to 1.68 times. Comparison with literature values demonstrates good agreement across the M dwarf temperature range (3000-4000 K), with a small systematic offset of 12 K and a scatter of 70 K, improving to 47 K for high S/N spectra. Surface gravity measurements show a minimal bias of $-$0.04 dex and a scatter of 0.17 dex, which improves to 0.12 dex for high quality data. The metallicity comparison reveals a slight systematic offset of $-$0.06 dex and scatter of 0.20 dex, improving to 0.14 dex for S/N$>$50 spectra. These results are further supported by comparison with APOGEE DR17, showing consistent precision levels (temperature scatter of 67 K and metallicity scatter of 0.17 dex). Further validation using FGK+M wide binaries shows metallicity differences with a scatter of 0.24 dex for the full sample (S/N$>$5), improving to 0.21 dex at S/N$>$20, and reaching 0.15 dex for high S/N spectra while maintaining broad metallicity coverage.

The resulting catalog provides comprehensive characterization of 507,513 M dwarf stars, including fundamental parameters, radial velocities, and activity indicators. We employed multiple classification methods and included flags for variability and binarity through cross-matching with various catalogs. This represents one of the largest uniformly analyzed samples of M dwarf stars, offering valuable insights into the Galaxy's most numerous stellar population. The complete dataset is publicly available, providing a rich resource for future studies of stellar and Galactic evolution.

\begin{acknowledgments}
\nolinenumbers
ZS acknowledges Jia-Dong Li and Yu-Ming Fu for very helpful discussions. This work was funded by the National Natural Science Foundation of China (NSFC Grant No.12090040,12090044,12173021,12133005,12103068). Guoshoujing Telescope (the Large Sky Area Multi-Object Fiber Spectroscopic Telescope, LAMOST) is a National Major Scientific Project built by the Chinese Academy of Sciences. Funding for the project has been provided by the National Development and Reform Commission. LAMOST is operated and managed by the National Astronomical Observatories, Chinese Academy of Sciences.
\end{acknowledgments}

\software{astropy \citep{2013A&A...558A..33A,2018AJ....156..123A}, 
		matplotlib \citep{2005ASPC..347...91B},
		pandas\citep{mckinney2010data},
		numpy \citep{4160250},
		scipy \citep{2019arXiv190710121V}
            sklearn \citep{scikit-learn}
        }

\bibliography{paper}{}
\bibliographystyle{aasjournal}

\end{CJK*}
\end{document}